\documentclass{aa}  

\usepackage{graphicx}
\usepackage{subcaption}
\usepackage{txfonts}
\usepackage{bm}
\usepackage{xcolor}

\usepackage{mathtools}
\usepackage{empheq}
\usepackage[colorlinks=true,linkcolor=blue,citecolor=blue,urlcolor=blue]{hyperref}
\usepackage{tikz}
\usepackage[utf8]{inputenc}

\def\ii{\textrm{i}}

\def\dd{\mathrm{d}}
\def\Ec{E_{\rm c}}
\def\Lm{\mathbb{L}_m}

\begin{document}

   \title{Nonlinear evolution of unstable solar inertial modes:\\ The case of  viscous modes on a differentially rotating sphere}

 \titlerunning{Amplitude saturation of high-latitude toroidal inertial modes} 
 
\author{
Muneeb~Mushtaq\inst{1} 
\and Damien~Fournier \inst{1} 
\and Rama~Ayoub\inst{4} 
\and Peter J. Schmid\inst{4} 
\and Laurent~Gizon\inst{1,2,3}\thanks{Corresponding author: \email{gizon@mps.mpg.de}}
}

\institute{\inst{1} Max-Planck-Institut f\"ur Sonnensystemforschung, Justus-von-Liebig-Weg 3, 37077 G\"ottingen, Germany \\
\inst{2} Institut f\"ur Astrophysik und Geophysik, Georg-August-Universit\"at G\"ottingen, Friedrich-Hund-Platz 1, 37077 G\"ottingen, Germany\\
\inst{3} Center for Astrophysics and Space Science, NYUAD Institute, New York University Abu Dhabi, Abu Dhabi, UAE \\
\inst{4} Department of Mechanical Engineering, King Abdullah University of Science and Technology, Thuwal, Saudi Arabia
}

\date{}

\abstract{
On the Sun, the inertial mode with the largest observed amplitude (rms velocity exceeding $10$~m/s) is the high-latitude mode with longitudinal wavenumber $m=1$. In two dimensions, linear theory predicts that this mode is unstable due to a shear instability associated with  latitudinal differential rotation (fast equator, slower polar regions).}{We investigate the evolution of this instability  numerically and theoretically.}{The nonlinear vorticity equation is solved using direct numerical simulations in the time domain. The only control parameter is the Ekman number $E$. For $10^{-3}\lesssim{E}<\Ec\approx1.5\times10^{-3}$, only the high-latitude $m=1$ mode is unstable. We extract its saturation amplitude as a function of $E$ and compare the results with predictions from two perturbative approaches in nonlinear stability theory.
}{
The simulations reveal a supercritical Hopf bifurcation. Near onset, the mode amplitude is well described by the Landau equation
$d|A|/dt=\sigma_I|A|+\beta_I|A|^3$,
derived from the perturbation theories, with a positive linear growth rate $\sigma_I$ and a negative nonlinear coefficient $\beta_I$. This description remains valid throughout the range $10^{-3}\lesssim{E}<\Ec$. The coefficient $\beta_I$ depends weakly on $E$, implying that the saturated amplitude scales approximately as $|A|\propto\sigma_I^{1/2}$. The equilibrium mode shape consists of the $m=1$ fundamental mode and higher harmonics $m=2$ and $m=3$, whose amplitudes scale approximately as $\sigma_I^{m/2}$. The saturation mechanism arises from the Reynolds stress that feeds back on and smoothes the latitudinal differential rotation. At $E=4\times10^{-4}$, consistent with solar-like values inferred from supergranular turbulent viscosity, the saturated velocity reaches $28$~m/s, comparable to the  observed value on the Sun.
}{
In two dimensions, the onset and saturation of the solar $m=1$ inertial mode are governed by a supercritical Hopf bifurcation associated with a shear instability, well described by weakly nonlinear theory. These results should not be overinterpreted, however, since in three-dimensional models the instability is baroclinic in nature, implying different underlying physics.
}

\keywords{Sun: helioseismology - Sun: oscillations - Methods: numerical - Instabilities}

   \maketitle

\section{Introduction}

Inertial modes are low-frequency global modes of oscillations driven by the Coriolis force in rotating fluids. Since the first discovery of Rossby modes on the Sun \citet{Loeptien2018}, several classes of modes, including high-latitude modes \citep{Gizon2021} and mode of rotating convection \citep[HFR,][]{Hanson2022, Hanasoge2026}, have been reported. Observations include frequencies, amplitudes, linewidths, and surface velocity eigenfunctions  \citep[][and references therein]{Gizon2024}. Eigenvalue solvers have been developed \citep{Fournier2022, Bekki2022b,  Bhattacharya2023, Mukhopadhyay2025} to identify the modes, and to study their linear stability.  Normal mode calculations, however, are insufficient to explain the mode amplitudes.

The amplitudes of the linearly stable modes, such as the Rossby modes,  have been studied using semi-analytical methods and nonlinear simulations. \citet{Philidet2023},  \citet{Bekki2022a}, and \citet{Blume2024} argue that these modes are stochastically excited and damped by near-surface turbulent convection, leading to amplitudes of order 1~m/s, consistent with the  observations.

\begin{figure*}[]  
\centering 
\begin{minipage}{0.45\textwidth}
    \centering
    \begin{tikzpicture}[xscale=-1] 
        \draw[-][red, dashed, thick] (2,0) -- (5.5,0);
        \draw[->] (5.4,-0.5) -- (5.4,4);

        \draw[->] (1.8,3.7) -- (1.8,0.15);
        \draw[->] (1.3,3.7) -- (1.3,0.15);
        \draw[->] (0.8,3.7) -- (0.8,0.15);
        \draw[->] (0.3,3.7) -- (0.3,0.15);
    
        \draw[->] (4.5,0.15) -- (4.5,2.4);
        \draw[->] (4,0.15) -- (4,2.2);
        \draw[->] (3.5,0.15) -- (3.5,2);
        \draw[->] (3,0.15) -- (3,1.8);
        \draw[->] (2.5,0.15) -- (2.5,1.2);
    
        \draw[->] (4.5,3.7) -- (4.5,2.9);
        \draw[->] (4,3.7) -- (4,2.7);
        \draw[->] (3.5,3.7) -- (3.5,2.5);
        \draw[->] (3,3.7) -- (3,2.3);
        \draw[->] (2.5,3.7) -- (2.5,2);

        \node at (5.6,-0.0) {$0$};
        \node at (-0.5,-0.0) {$E$};
        \node at (2,-0.25) {$\Ec$};
        \node at (5.85, 2) {$|A|$};  

        \draw[blue, thick, <-] (-0.3,0) -- (2,0) .. controls (2.4,2.4) and (3.5,2.5) .. (5,2.7);

        \filldraw[black] (2,0) circle (1.5pt);
    \end{tikzpicture}
\end{minipage}
\hspace{0.5cm}  
\begin{minipage}{0.45\textwidth}
    \centering
    \begin{tikzpicture}[xscale=-1]
        \draw[-][red, thick, dashed] (2,0) -- (5.5,0);
        \draw[->] (5.4,-0.5) -- (5.4,4);  

        \draw[->] (1,1) -- (1,0.15);
        \draw[->] (0.6,3.8) -- (0.6,0.15);
        \draw[->] (0.2,3.8) -- (0.2,0.15);
        
        \draw[->] (1.4,0.65) -- (1.4,0.15);
        \draw[->] (1.8,0.4) -- (1.8,0.15);
    
        \draw[->] (1.8,0.75) -- (1.8,1.9);
        \draw[->] (1.4,0.9) -- (1.4,1.7);

        \draw[->] (4.5,0.15) -- (4.5,3.2);
        \draw[->] (4,0.15) -- (4,3);
        \draw[->] (3.5,0.15) -- (3.5,2.8);
        \draw[->] (3,0.15) -- (3,2.6);
        \draw[->] (2.5,0.15) -- (2.5,2.4);
    
        \draw[->] (4.5,3.8) -- (4.5,3.5);
        \draw[->] (4,3.8) -- (4,3.3);
        \draw[->] (3.5,3.8) -- (3.5,3.1);
        \draw[->] (3,3.8) -- (3,2.9);
        \draw[->] (2.5,3.8) -- (2.5,2.7);
        \draw[->] (1.8,3.8) -- (1.8,2.3);
        \draw[->] (1.4,3.8) -- (1.4,2.1);
        \draw[->] (1.0,3.8) -- (1.0,1.7);

        \node at (5.6,-0.0) {$0$};
        \node at (-0.5,-0.0) {$E$};
        \node at (2,-0.25) {$\Ec$};
        \node at (5.85,2) {$|A|$};  

        \draw[blue, thick, <-] (-0.3,0) -- (2,0);
        \draw[red, thick, dashed] (2,0) -- (2,0) .. controls (2.13,0.6) and (1.2,0.7) .. (1,1.2);

        \draw[blue, thick] (1,1.2) .. controls (0.8,1.5) and (2.12,2.7) .. (5,3.5); 

        \filldraw[black] (2,0) circle (1.5pt);
    \end{tikzpicture}
\end{minipage}
\caption{Schematic diagram illustrating the two  bifurcation types:  supercritical (left) and subcritical (right). 
The thick blue curves show the stable equilibrium wave amplitudes as a function of Ekman number $E$, while the dashed red curves corresponds to the unstable solutions. The black arrows indicate the different possible evolutions of the   system, which can be computed numerically. We will consider two  perturbation methods to obtain the equilibrium amplitudes in the case of supercritical bifurcation, whereby the small parameter is either the distance to the critical Ekman number $E_{\rm c}$ or the  amplitude itself. } \label{fig:sup_sub_bif}  
\end{figure*}
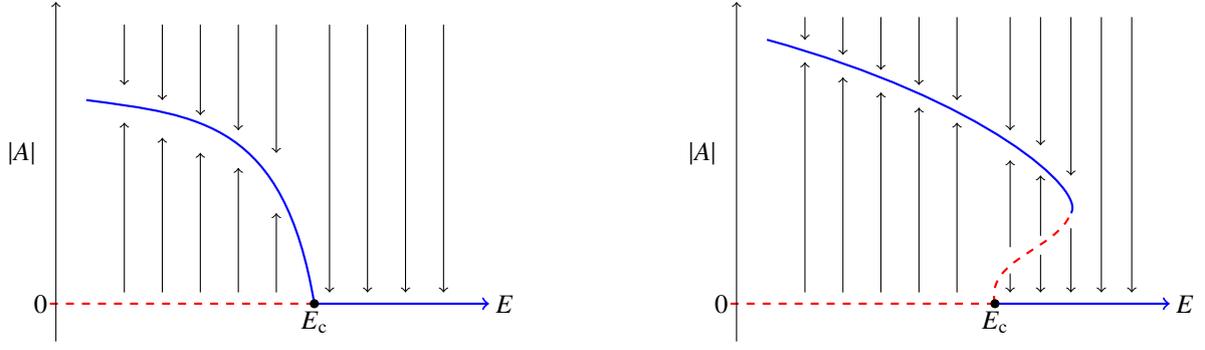

The high-latitude inertial modes have the largest amplitudes on the Sun. In particular, the $m=1$ high-latitude mode reaches amplitudes in the range $5$–$20$~m/s over the course of the solar cycle \citep{Liang2025,Lekshmi2026}, with an average value of $\sim 10$~m/s. Eigenvalue calculations in two horizontal dimensions \citep{Fournier2022} and in  three-dimensional models \citep{Bekki2022b} indicate that these modes are linearly unstable.
Numerical simulations of solar convection \citep{Bekki2024,SouzaGomes2025} show that the $m=1$ high-latitude mode is baroclinically unstable and  back-reacts on the differential rotation via equatorward transport of entropy and the thermal wind balance. In this scenario, this high-latitude mode plays a key role controlling the maximum allowed differential rotation. 
Such nonlinear simulations capture the full dynamics, including mode-mode interactions, kinetic and heat transfer mechanisms, but are computationally expensive and non-trivial to interpret. Moreover, it is challenging to obtain a solar-like differential rotation from first principles in these simulations --- a key ingredient to model the high-latitude inertial modes, which are very sensitive to the values of the latitudinal shear.

In this study, we compute the linear growth rate and the nonlinear saturation amplitude of the high-latitude $m=1$ unstable toroidal mode on a differentially rotating sphere, extending the linear analysis of \citet{Fournier2022}. In this  system, angular momentum is redistributed solely through the Reynolds stresses associated with the inertial modes, representing a significant simplification over the full dynamics in three dimensions. Nevertheless, this model provides valuable insight into the hydrodynamical interactions between the fundamental mode (the most unstable linear mode) and the base flow. It also allows us to compute the equilibrium state, which includes higher harmonics\footnote{By convention, the $m$-th harmonic denotes the oscillation with frequency $m$ times that of the fundamental mode, in accordance with, for example, \citet{Schmid2012}.}  with $m\ge 2$. In the literature \citep[see, for instance,][]{Bagheri2013}, these nonlinear states (fundamental and higher harmonics) also referred as the asymptotic Koopman modes .

Furthermore, the two-dimensional nonlinear problem is more amenable to an application of
weakly nonlinear (WNL) theory, which offers a promising alternative to fully nonlinear numerical solvers under certain conditions. The term "weakly" reflects the fact that the analysis does not account for the full extent of nonlinear effects but instead incorporates only a minimal set of nonlinear interactions  necessary to reach saturation. A central result of WNL theory is a time-dependent amplitude equation, originally introduced in fluid dynamics by \citet{Landau1944}, 
\begin{equation}
    \frac{\mathrm{d}|A|}{\mathrm{d}t} = \sigma_I |A| + \beta_I |A|^3 + \mathcal{O}(|A|^5) , \label{eqn:amplitude_equation}
\end{equation}
where $\sigma_I$, the first Landau coefficient, represent the linear growth rate of the mode ($\sigma_I>0$ for an unstable mode) and $\beta_I$ is known as the second Landau coefficient. If $\beta_I < 0$, the system is called supercritical: any disturbance beyond criticality will bring the system to the same final state as illustrated in the left panel of Fig.~\ref{fig:sup_sub_bif}. When there is no unstable mode (i.e., the Ekman number is larger than the critical value $\Ec$), any disturbance will decay in a finite amount of time and the system will return to its initial (trivial equilibrium) state. For $E < \Ec$, any disturbance will bring the system to a new (nontrivial) equilibrium state. If $\beta_I > 0$, the bifurcation is subcritical and the final state depends on the strength of the perturbation (right panel of Fig.~\ref{fig:sup_sub_bif}). In this case, higher-order terms in the amplitude equation are required to determine the equilibrium amplitude. 

The amplitude equation, Eq.~(\ref{eqn:amplitude_equation}) was first formally derived by \citet{Stuart1960}  for the Orr–Sommerfeld equation in the case of a plane Poiseuille flow, using  a multiscale expansion. 
Related problems for different shear flows in the $\beta$-plane where studied by, for example,  \citet{Burns1983} and \citet{Churilov1985}. 
Remarkably, our problem also consists of solving a  modified Orr–Sommerfeld equation, where the latitudinal differential rotation plays the role of the shear flow \citep{Gizon2020,Fournier2022}. These considerations motivate the application of weakly nonlinear theory in the supercritical regime. 

In the following, we show numerically that the $m=1$ high-latitude mode is the most unstable mode, that it undergoes a supercritical bifurcation, and that weakly nonlinear theory captures its nonlinear evolution near onset (for Ekman numbers not too far from the  solar values).

\section{Setup of the problem}

\subsection{Governing equation}

We consider the (nonlinear) propagation of purely toroidal modes on a two-dimensional differentially rotating sphere of radius $r$. While highly-simplified, this problem contains differential rotation and viscosity, two essential ingredients to study the different families of inertial modes \citep{Fournier2022}. 

In a frame rotating at $\Omega_\mathrm{ref}$, the momentum equation for incompressible viscous flow $\bm{u}$ is 
\begin{equation}
    \frac{\partial \bm{u}}{\partial t} + (\bm{u} \cdot \mathbf{\nabla})\bm{u} + 2 \, \Omega_\mathrm{ref} \hat{\bm{z}} \times \bm{u} = -\nabla\Pi + \nabla \cdot \bm{\mathcal{D}} , 
\label{eqn:Navier_Stokes1} 
\end{equation}
where $\Pi$ is a potential from which all the conservative forces (such as pressure and gravity) are derived, and $\bm{\mathcal{D}}$ is the viscous stress tensor. We have used $\Omega_\mathrm{ref}/2\pi = 456.03$ nHz in our present analysis. We work in spherical coordinates $(r,\theta,\phi)$, where $\theta$ is the colatitude and $\phi$ is the longitude, with unit vectors $(\hat{\bm{r}},\hat{\bm{\theta}},\hat{\bm{\phi}})$. Assuming purely toroidal motions, the flow can be written in terms of the stream function $\Psi$ such that 
\begin{equation}
    \bm{u} = \frac{1}{r \sin \theta}\frac{\partial \Psi}{\partial \phi}\hat{\boldsymbol{\theta}} - \frac{1}{r}\frac{\partial \Psi}{\partial \theta} \hat{\boldsymbol{\phi}} .
\label{eqn:straamfun}
\end{equation}
Also, the radial vorticity $Z$ is defined as
\begin{equation}
    Z  := (\nabla \times \bm{u})\cdot \hat{\bm{r}} = -\Delta_\mathrm{h} \Psi ,
\label{eqn:vorticity}
\end{equation}
where $\Delta_{\rm h}$ denote the horizontal Laplacian
\begin{equation}
    \Delta_\mathrm{h}  = \frac{1}{r^2 \sin \theta}\frac{\partial}{\partial \theta}\left(\sin \theta \frac{\partial }{\partial \theta}\right) + \frac{1}{r^2 \sin^2 \theta}\frac{\partial^2 }{\partial \phi^2} .
\label{eqn:nablah}
\end{equation}
Taking the radial component of the curl of Eq.~(\ref{eqn:Navier_Stokes1}), we obtain
\begin{equation}
   \frac{\partial Z}{\partial t} - \frac{2\Omega_\mathrm{ref}}{r^2} \frac{\partial \Psi}{\partial \phi} - \left[\nabla\times \left(\nabla \cdot \bm{\mathcal{D}}\right)\right]\cdot \hat{\bm{r}} = J[Z, \Psi] ,
\label{eqn:streamNL}
\end{equation}
where $J[Z, \Psi]$ is the (Jacobian) nonlinear term 
\begin{equation}
    J[Z, \Psi] = \frac{1}{r^2\sin\theta} \left(\frac{\partial \Psi}{\partial \theta} \frac{\partial Z}{\partial \phi}  -  \frac{\partial \Psi}{\partial \phi} \frac{\partial Z}{\partial \theta}\right) .  
\label{eqn:sph_jacobian}
\end{equation}
It remains to discuss the viscous stress tensor, which contains a dissipative ($\bm{\mathcal{D}}^\nu$) and conservative ($\bm{\mathcal{D}}^\Lambda$) parts \citep[see, for instance,][]{Rudiger1989}:
\begin{equation}
    \mathcal{D}_{ij} = \mathcal{D}^\nu_{ij} + \mathcal{D}^\Lambda_{ij} = \nu \left( S_{ij} + \Lambda_{ij} \Omega \right) , 
\label{eqn:viscous_tensor}
\end{equation}
where $\bm{S} = \left[\nabla\bm{u} + (\nabla \bm{u}){^T}\right]/2$ is the strain tensor and $\nu$  is the eddy or turbulent viscosity. The angular velocity, $\Omega$, is defined in terms of the axisymmetric  component of the azimuthal velocity, 
\begin{equation}
    \Omega(t,\theta) = \Omega_\mathrm{ref} + \frac{1}{r\sin\theta} \frac{1}{2\pi} \int_0^{2\pi} u_\phi(t,\theta,\phi) \dd \phi .
\label{eqn:rot_evol}
\end{equation}
The unit-less tensor elements $\Lambda_{ij}$  account for the non-diffusive angular momentum transport (the $\Lambda$ effect) which maintains a solar-like differential rotation. The divergence of the viscous stress tensor is 
\begin{equation}
    \nabla \cdot \bm{\mathcal{D}} = \nu \Delta \bm{u} +  \frac{\nu}{r\sin^2\theta} \frac{\partial \chi}{\partial \theta} \hat{\bm{\phi}} \quad \text{with} \quad \chi := \sin^2\theta \Lambda_{\theta\phi} \Omega . 
\label{eqn:div_visc_simp}
\end{equation}
where $\Delta  := \Delta_\mathrm{h} + 2/r^2$. The   term $2/r^2$ stem from taking the divergence of the strain tensor $\bm{S}$ in spherical geometry \citep[see, for instance,][their Appendix B]{Lindborg2022}; it was missing in our previous study \citep{Fournier2022}, but is necessary to ensure the conservation of angular momentum. Equation~\eqref{eqn:streamNL} leads to an equation for the radial vorticity $Z$:
\begin{equation}
    \frac{\partial Z}{\partial t} - \frac{2\Omega_\mathrm{ref}}{r^2}\frac{\partial \Psi}{\partial \phi} - \nu \Delta Z - \frac{\nu}{r^2\sin\theta} \frac{\partial}{\partial\theta} \left(\frac{1}{\sin\theta}  \frac{\partial \chi}{\partial\theta}\right) = J[Z, \Psi] . 
\label{eq:vorticity_nl}
\end{equation}

\subsection{Background differential rotation and the $\Lambda$ effect}

\begin{figure}[]
    \centering
    \includegraphics[width=0.8\linewidth]{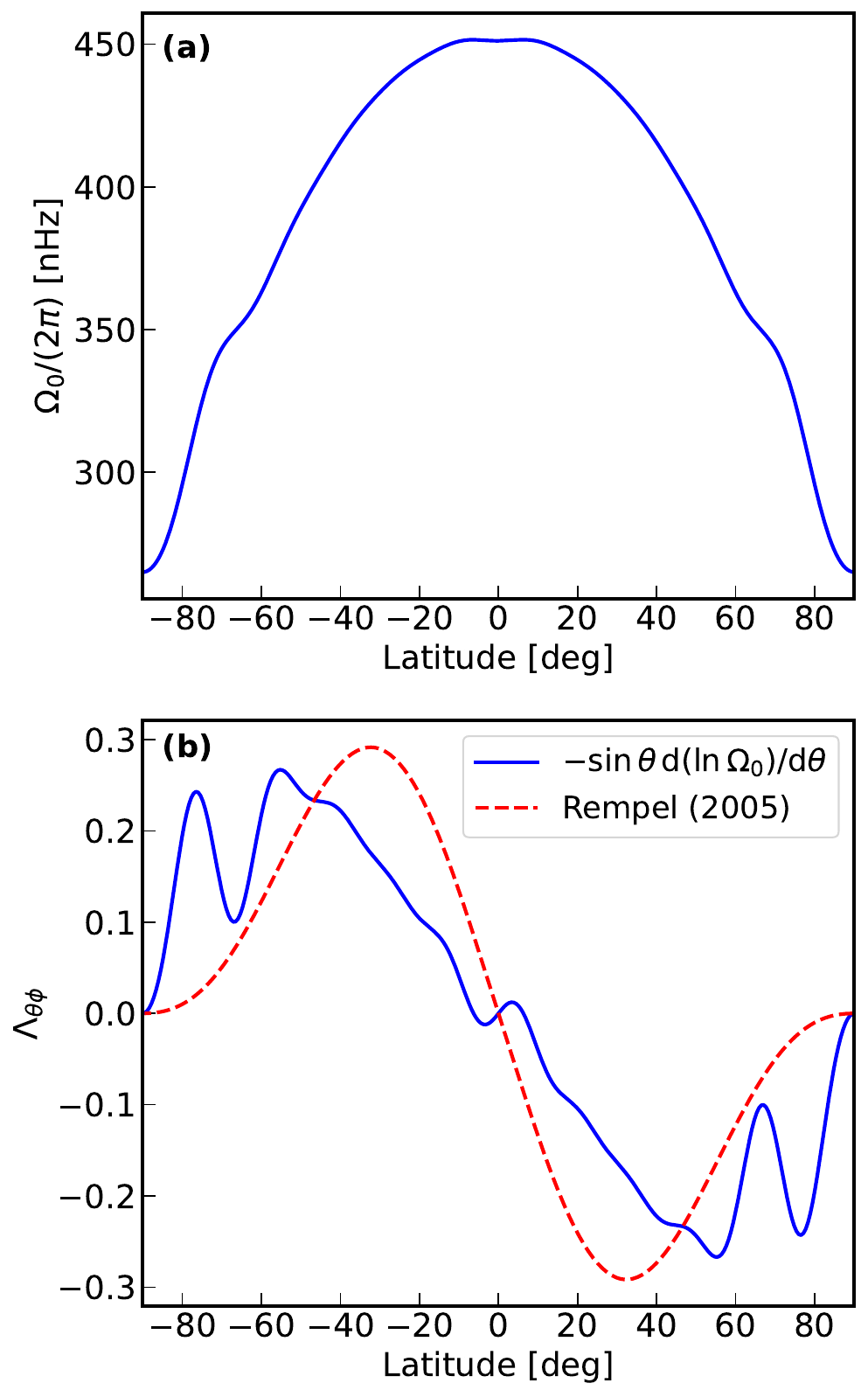}
\caption{Initial rotation and associated $\Lambda$-effect. (a) The surface solar differential rotation profile $\Omega_0(\theta)$ inferred by helioseismology from six years of HMI data  \citep[with 36~$a$-coefficients,][]{Larson2018}. (b) The corresponding  function $\Lambda_{\theta\phi}(\theta)$ computed using Eq.~(\ref{eqn:omega_ode}). For comparison, we overplot  the $\Lambda$-effect prescribed by \citet{Rempel2005}, that is $\Lambda_{\theta\phi} = - 0.8  \sin^2 \theta \cos \theta \sin\left(\theta + 15^\circ \right)$ in the northern hemisphere.}
\label{fig:rot_lambda}
\end{figure}

Equation~\eqref{eq:vorticity_nl} has to be complemented by an initial condition. We initialize our simulation with a solar-like differential rotation profile $\Omega_0(\theta)$ and add a small amount of random noise to excite the system. We consider a differential rotation profile at the Sun's surface $(r=R_\odot)$ inferred from six years of HMI data with 36 Helioseismic $a$-coefficients \citep[][see Fig.~\ref{fig:rot_lambda}a]{Larson2018}.
The $\Lambda$-effect is chosen such that Eq.~\eqref{eq:vorticity_nl} is satisfied at time $t=0$ which implies (see Appendix~\ref{app:lambda_effect_der} for the derivation)
\begin{equation}
    \Lambda_{\theta\phi} = -\sin \theta  \frac{\mathrm{d} \ln \Omega_0}{\mathrm{d } \theta} .
\label{eqn:omega_ode}
\end{equation}
The corresponding $\Lambda$ effect, $\Lambda_{\theta\phi}(\theta)$, computed using solar-like profile $\Omega_0(\theta)$, is shown in Fig.~\ref{fig:rot_lambda}b.   

\subsection{Equation for nonlinear waves}

We now consider the deviation from the initial (base) flow $\bm{u}_0 = (\Omega_0 - \Omega_\mathrm{ref})r \sin\theta\ \hat{\bm{\phi}}$ by writing the flow $\bm{u} = \bm{u}_0 + \bm{u}'$. The zeroth-order equation is given as
\begin{equation}
     \left(\Delta_{\rm h} + \frac{2}{r^2} \right) Z_0
    + \frac{1}{r^2\sin\theta} \frac{\partial}{\partial\theta} \left(\frac{1}{\sin\theta}  \frac{\partial}{\partial\theta} \left(\sin^2\theta \Lambda_{\theta\phi} \Omega_0 \right) \right) = 0 ,
\label{eqn:Z0}
\end{equation}
where 
\begin{equation}
    Z_0 = \left(\nabla \times \bm{u}_0\right)\cdot \hat{\bm{r}} = \frac{1}{\sin\theta} \frac{\dd}{\dd \theta} \left( \sin^2\theta \, (\Omega_0 - \Omega_\mathrm{ref}) \right) ,
\end{equation}
is the initial radial vorticity. Equation (\ref{eqn:Z0}) is satisfied with our choice of the $\Lambda$ effect from the previous section.

In the present analysis, we did not consider the modification of $\Lambda$ due to a change in rotation; that is, we suppose that $\Lambda$ is constant throughout the simulation. This assumption is often used in numerical simulations of the solar convection zone \citep[see, for instance,][]{Rempel2005,Bekki2022a} but could be omitted if necessary. In that case, the vorticity equation~\eqref{eq:vorticity_nl} can be written in terms of the perturbed vorticity $Z' = Z - Z_0$ and stream function $\Psi' = \Psi - \Psi_0$
\begin{equation}
    \left(\frac{D}{D t} - \nu \Delta\right)Z' + \frac{1}{r^2\sin\theta}\frac{\partial \tilde{Z}_0}{\partial \theta} \frac{\partial \Psi'}{\partial \phi} = J[Z', \Psi'] , 
\label{eq:vorticity_nl2}
\end{equation}
where $\tilde{Z}_0 = 1/\sin\theta \, \partial_\theta(\sin^2\theta \, \Omega_0)$ and $D_t = \partial_t + (\Omega_0 - \Omega_\mathrm{ ref})\partial_\phi$ stands for the material derivative in the rotating reference frame.

\section{Direct numerical simulations (DNS)}

\subsection{Dimensionless form of the equation of motion}

We introduce the dimensionless quantities
\begin{equation}
    \hat{Z} = \frac{Z'}{\Omega_\mathrm{ref}}, \quad  \hat{\Psi} = \frac{\Psi'}{r^2 \Omega_\mathrm{ref}}, \quad  \hat{Z}_0 = \frac{\tilde{Z}_0}{\Omega_\mathrm{ref}}, \quad \hat{t} = \Omega_\mathrm{ref} t ,
\label{eqn:scalings}
\end{equation}
and operators $\hat{\Delta} = r^2 \Delta$ and $\hat{J} = r^2 J$. The normalized radial vorticity and the stream function are related through $\hat{Z} = -\hat{\Delta}_{\rm h}\hat{\Psi}$. Defining the scaled differential rotation $\delta(\theta)$ and the Ekman number $E$ as 
\begin{equation}
    \delta(\theta) := \frac{\Omega_0(\theta) - \Omega_\mathrm{ ref}}{\Omega_\mathrm{ref}} \, , \quad \text{and} \quad E := \frac{\nu}{r^2 \Omega_\mathrm{ref}},
\end{equation}
Eq.~\eqref{eq:vorticity_nl2} can be written in dimensionless form as 
\begin{equation}
    \frac{\partial \hat{Z}}{\partial \hat{t}}  + \delta \frac{\partial \hat{Z}}{\partial \phi} + \frac{1}{\sin\theta} \frac{\partial \hat{Z}_0}{\partial \theta} \frac{\partial\hat{\Psi}}{\partial \phi}  - E \hat{\Delta}  \hat{Z}
    =  \hat{J}[\hat{Z}, \hat{\Psi}] . 
\label{eqn:forward_norm}
\end{equation}
The total flow ${\bm{u}}$ can be obtained from  $\hat{\Psi}$ according to
\begin{equation}
    \bm{u} = r\sin\theta \, (\Omega_0(\theta)-\Omega_\mathrm{ref}) \hat{\bm{\phi}} 
    +  r \, \Omega_\mathrm{ref} \left(\hat{\bm{\theta}} \frac{1}{\sin \theta} \frac{\partial }{\partial \phi} - \hat{\bm{\phi}} \frac{\partial}{\partial \theta} \right)\hat{\Psi},
\label{eqn:straamfun1}
\end{equation}
where the first term on the right-hand side represents the background flow $\bm{u}_0$, and the second term corresponds to the perturbation $\bm{u}'$ around this base flow.

\subsection{Temporal scheme}

The temporal evolution of the radial vorticity field described by Eq.~\eqref{eqn:forward_norm} can be written as
\begin{equation}
    \frac{\partial \hat{Z}}{\partial \hat{t}} = F[\hat{Z}] , \label{eqn:psi_eqn}
\end{equation}
where $F[\hat{Z}] = \mathcal{L}[\hat{Z}] + \mathcal{N}[\hat{\Psi},\hat{\Psi}]$ with $\hat{Z} = -\Delta_{\rm h} \hat{\Psi}$ and the linear operator $\mathcal{L}$ and the bilinear form $\mathcal{N}$ are given by
\begin{align}
    \mathcal{L}[\hat{Z}] &:= -\delta \frac{\partial \hat{Z}}{\partial \phi} - \frac{1}{\sin\theta} \frac{\partial \hat{Z_0}}{\partial \theta} \frac{\partial\hat{\Psi}}{\partial \phi}  + E \hat{\Delta}  \hat{Z}, \label{eq:linear_part_pde} \\
    \mathcal{N}[\hat{\Psi}_1, \hat{\Psi}_2] &:= \hat{J}[-\Delta_{\rm h} \hat{\Psi}_1, \hat{\Psi}_2] \label{eq:non-linear_part_pde}.
\end{align}
For the time discretization, we choose an explicit scheme as the Courant-Friedrichs-Lewy (CFL) condition is not too restrictive. In particular, we use the Adams–Bashforth scheme  of order three \citep[see e.g.,][]{Durran1991} such that
\begin{equation}
    \hat{Z}_{n+1} = \hat{Z}_n + \frac{\Delta t}{12} \left( 23 F[\hat{Z}_n] - 16 F[\hat{Z}_{n-1}] + 5 F[\hat{Z}_{n-2}] \right) ,
\end{equation}
where $\hat{Z}_n = \hat{Z}(\hat{t} = n \, \Delta t)$. The time step $\Delta t$ is chosen so that the CFL condition is respected and usually amounts to a few hours. $\hat{Z}_0$ is prescribed through the initial condition. For the first iterations, we use a scheme of order one to obtain $\hat{Z}_1$ and of order two to obtain $\hat{Z}_2$.

\subsection{Spatial discretization}

At a given time step, $\hat{Z}_n(\theta,\phi)$ is decomposed into spherical harmonics up to a maximum degree $L_{\rm max}$ ($L_{\rm max} = 200$ in this study)
\begin{equation}
    \hat{Z}_n(\theta,\phi) = \sum_{\ell=0}^{L_{\rm max}} \sum_{m=-\ell}^\ell \hat{Z}_n^{\ell m} Y_\ell^m(\theta,\phi) .
\end{equation}
We need to evaluate the spherical harmonic coefficients of $F[\hat{Z}_n]$.
The linear part $\mathcal{L}[\hat{Z}_n]$ defined by Eq.~\eqref{eq:linear_part_pde} can be computed directly in spherical harmonic space from $\hat{Z}_n^{\ell m}$ and the associated stream function coefficients $\hat{\Psi}_n^{\ell m} = \hat{Z}_n^{\ell m} / (\ell(\ell+1))$ in a similar manner as in \citet{Fournier2022}. For the nonlinear part, we reconstruct the $\theta-$ and $\phi-$derivatives of $\hat{\Psi}$ and $\hat{Z}$ from the properties of spherical harmonics using the \texttt{shtns} library \citep{shtns}. We can then build the Jacobian term and project it onto spherical harmonics.

\section{Numerical results}

The simulations are initialized with the surface solar differential rotation corresponding to the profile from \citet{Larson2018} obtained with 36 $a$-coefficients from six years of HMI data. On top of the rotation, we add white noise (in spherical harmonic space) to destabilize the system. We then run the simulations until  they reach an equilibrium state. 

\begin{figure}[t]
  \centering
    \centering
    \includegraphics[width=0.9\linewidth]{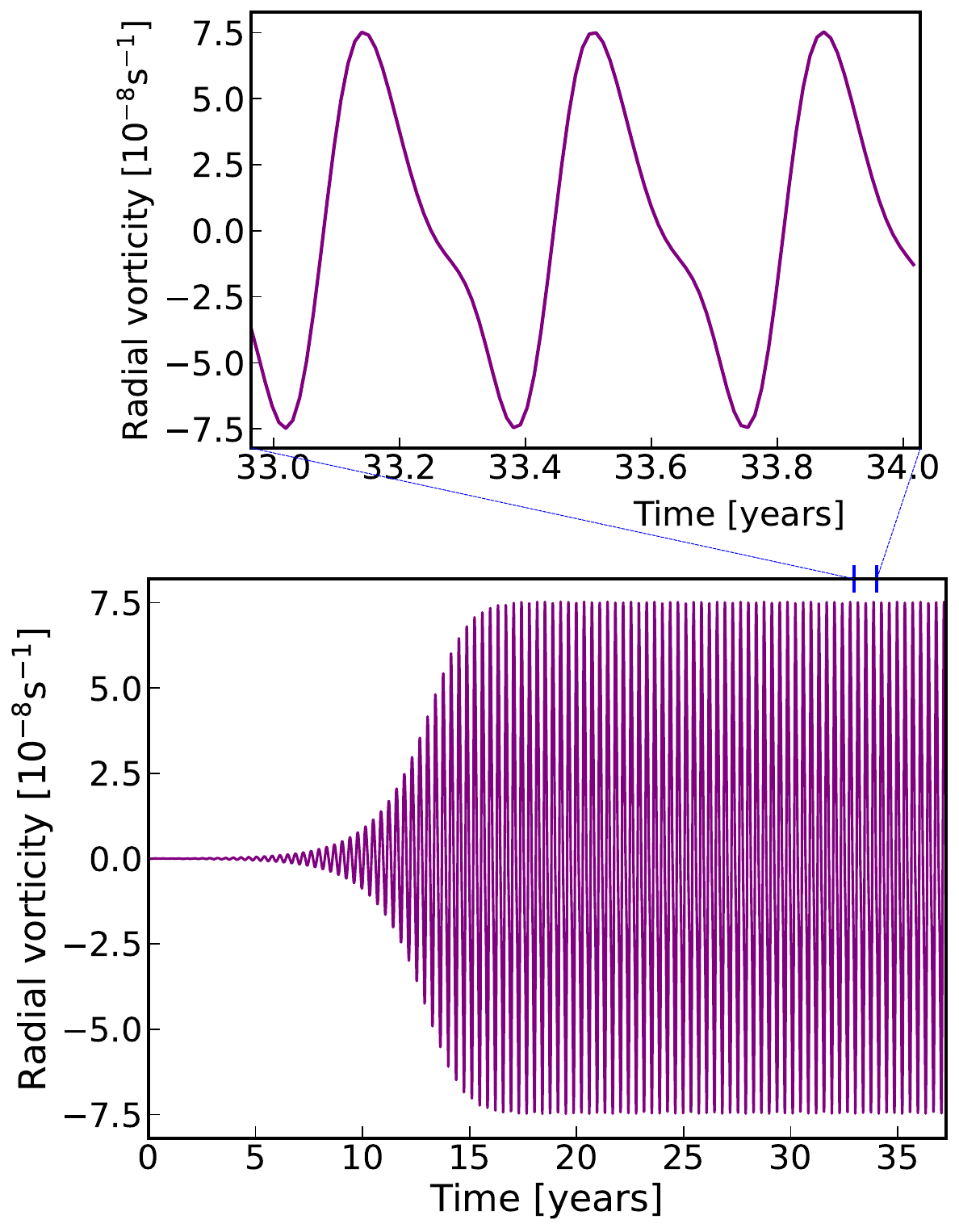}
  \caption{Time evolution of the radial vorticity $Z(t, \theta, \phi) - \langle Z(t,\theta,\phi) \rangle_\phi$ at a fixed point $(\theta, \phi) = (60^\circ, 30^\circ)$ from DNS for an Ekman number, $E = 10^{-3}$. The upper panel is a zoom between the two blue vertical lines on the bottom panel to show that the oscillations do not consist of a single harmonic due to nonlinearities in the system.} 
\label{fig:unfiltered_vorticity}
\end{figure}
\begin{figure*}[ht]
  \centering
    \includegraphics[width=\linewidth]{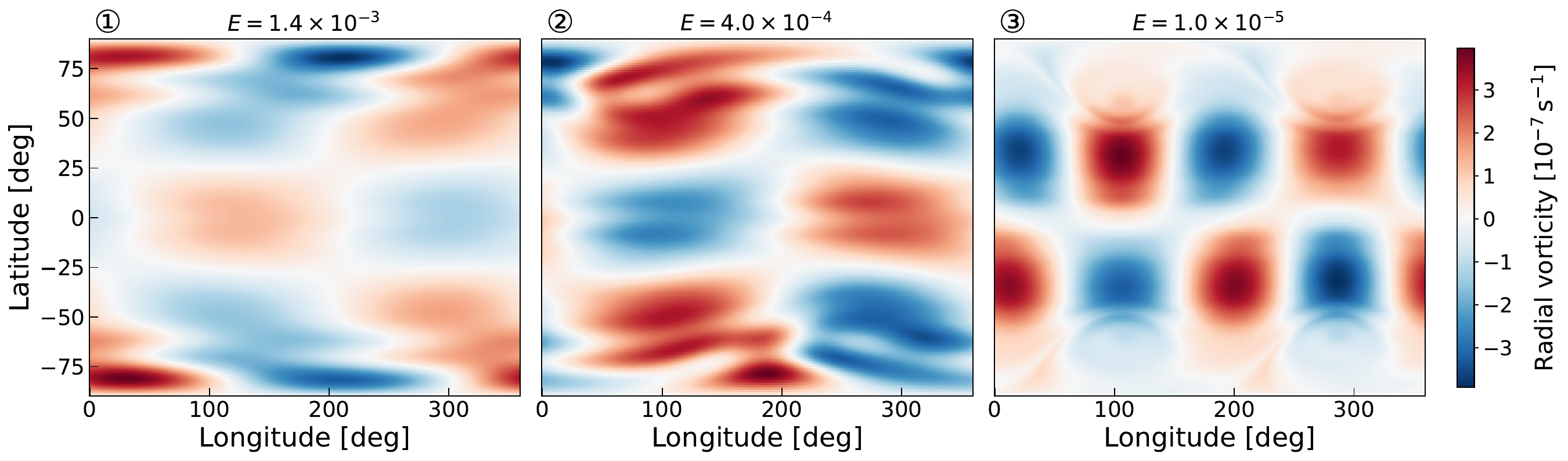} 
  \caption{Snapshots of the two-dimensional radial vorticity $Z(t, \theta, \phi) - \langle Z(t, \theta, \phi)\rangle_\phi$ at time $t =120 $ year for three different Ekman numbers, $E = 1.4 \times 10^{-3}, 4 \times 10^{-4},$ and $10^{-5}$. The time variations can be seen on the \href{https://doi.org/10.17617/3.U2TWXU}{online movie}.} 
\label{fig:snapshot_vorticity}
\end{figure*}

\subsection{Time evolution and saturation regime}

Figure~\ref{fig:unfiltered_vorticity} shows the radial vorticity as a function of time at a specific location on the sphere, $(\theta, \phi) = (60^\circ, 30^\circ)$, for an Ekman number of $E = 10^{-3}$. In this case, only one ($m = 1$) mode is linearly unstable, and we observe three main phases in time:
\begin{itemize}
    \item Initial transient phase: a highly nonlinear phase that depends on the initial condition. The initial small structures are organizing to trigger an instability.
    \item Amplification phase: the unstable mode is growing in amplitude until it sufficiently interacts with the background flow to stabilize the system.
    \item Saturation regime: the system is stable and has reached an equilibrium state. The signal oscillates at the frequency of the most dominant mode ($m=1$ mode with a frequency of $\approx -87$~nHz) but the time dependence is not exactly harmonic due to nonlinearities.
\end{itemize}

Snapshots in the equilibrium phase of three simulations at different Ekman numbers are presented in Fig.~\ref{fig:snapshot_vorticity}. They represent the radial vorticity minus the longitudinally-averaged vorticity for better visibility. For large Ekman numbers ($E = 1.4 \times 10^{-3}$), only the $m=1$ mode is unstable and the solution has mostly an $m=1$ component at the end of the simulation. 
Decreasing viscosity, several modes are unstable, and the dominant pattern at the end of the simulation is still the $m=1$ for $E = 4 \times 10^{-4}$ but becomes an $m=2$ for smaller Ekman numbers. For very small viscosity ($E = 10^{-5}$), the cat's eye corresponding to the Kelvin-Helmholtz instability become visible, and the eigenfunction differs significantly from the linear eigenfunction \citep[see Fig.~C.7 in][]{Fournier2022}.

\subsection{Supercritical (Hopf) bifurcation}

\begin{figure}[t]
    \centering
    \includegraphics[width=0.9\linewidth]{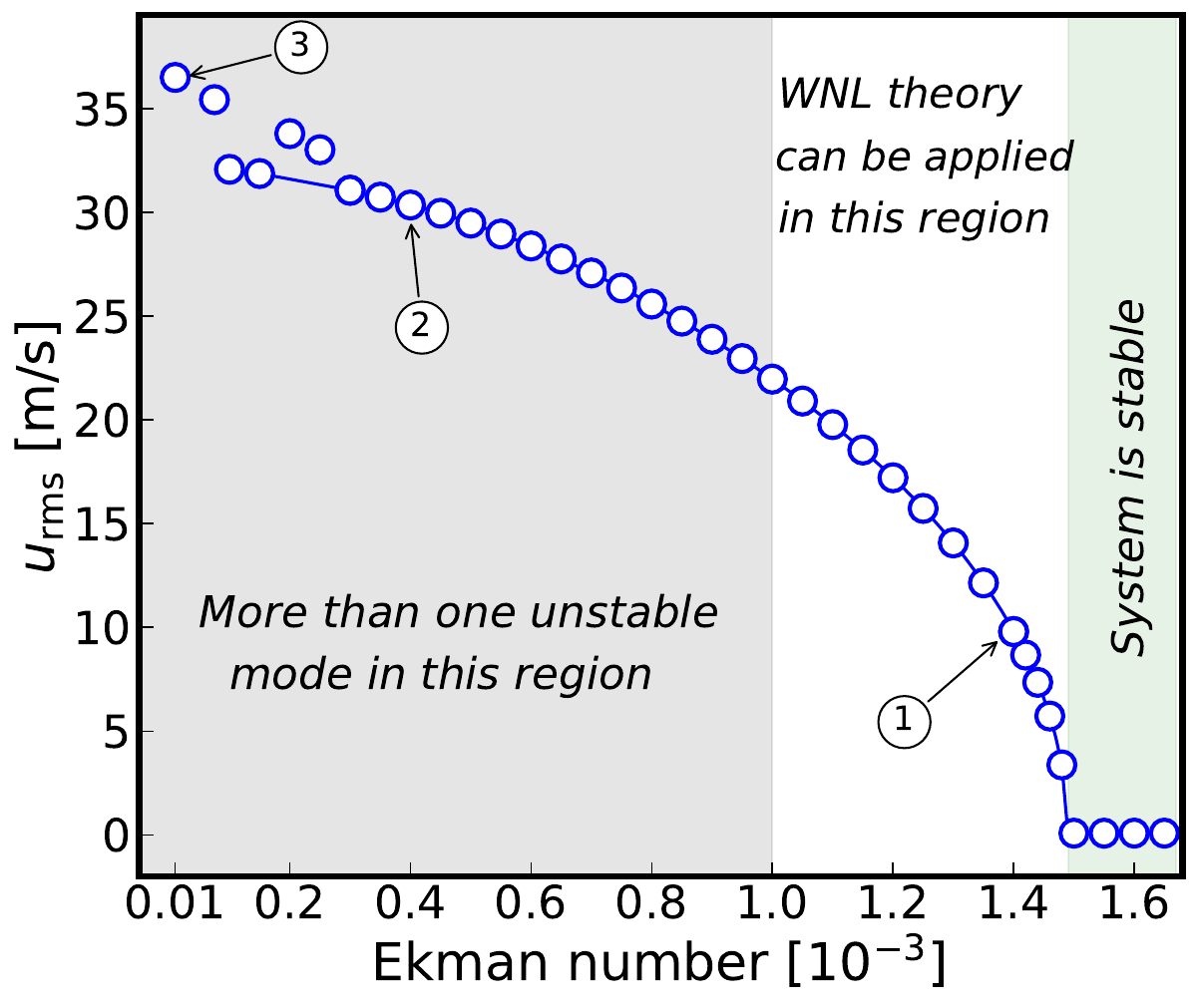}
\caption{RMS velocity (Eq.~\eqref{eqn:urms_amp}) obtained using the unfiltered velocity fields,  as a function of the Ekman number. The green shaded region corresponds to $E > \Ec$ where no modes are unstable. The white region corresponds to the region where only one mode is unstable, and hence the WNL theory can be applied, in contrast to the gray region. The points labeled as 1, 2, and 3, respectively, correspond to the Ekman numbers from the left, middle, and right subplots in Fig.~\ref{fig:snapshot_vorticity}.}
\label{fig:bifurcation_num}
\end{figure}

Figure~\ref{fig:unfiltered_vorticity} shows the evolution of the radial vorticity for a representative Ekman number, $E = 10^{-3}$. We performed simulations at different Ekman numbers and observed that all of them converged to a final steady state, independent of the strength of the initial perturbation, a characteristic of a supercritical bifurcation structure. Figure~\ref{fig:bifurcation_num} shows the final root-mean-square (RMS) velocity
\begin{equation}
    u^2_\mathrm{rms}(t) := \frac{1}{4 \pi}\int_0^{2\pi}  \int_0^{\pi} \left( u_\theta^2(t, \theta, \phi) + u_\phi^2(t,\theta,\phi) \right) \sin\theta \dd\theta \dd\phi ,
\label{eqn:urms_amp}
\end{equation}
as a function of the Ekman number.
For $E > \Ec$, all modes are linearly stable, and the system returns to its initial (trivial equilibrium) state. For $E < \Ec$, at least one mode is unstable and an instability develops. The saturation amplitude of the mode is increasing with the distance from the critical Ekman number. In Sect.~\ref{sec:weakly_nonlinear}, we present a WNL theory to explain the final amplitudes. This theory is developed when only one mode is unstable which is valid for $E \in [10^{-3}, 1.5 \times 10^{-3})$ in this setup. However, this theoretical framework can be extended, if multiple modes are unstable \citep[see, for example,][]{Dey2019, Cudby2021}. If too many modes are unstable, other numerical approaches, such as the Dynamic Mode Decomposition  \citep{Schmid2022}, are preferred.

\begin{figure*}[t]
    \centering
    \includegraphics[width=\linewidth]{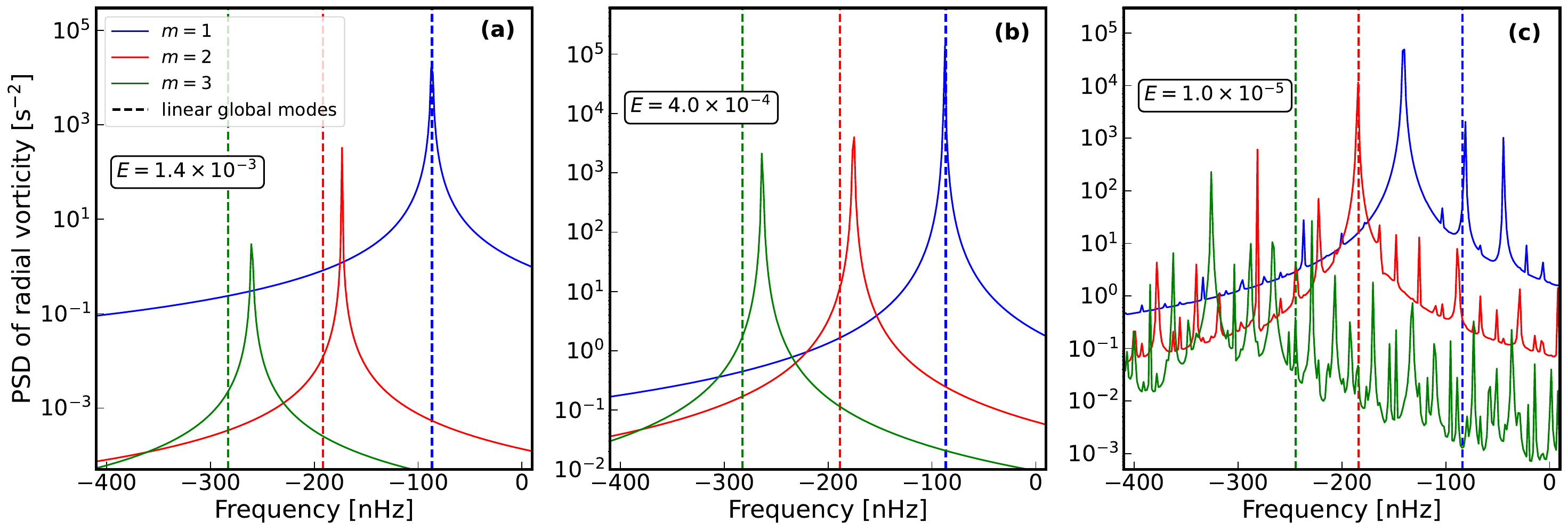}
\caption{PSD of the radial vorticity field $Z_m(t, \theta)$ obtained from numerical simulations for different azimuthal wavenumbers $m$ at Ekman numbers; (a) $E = 1.4 \times 10^{-3}$, (b) $E = 4.0 \times 10^{-4}$ and (c) $E = 10^{-5}$. The solid curves are from the DNS and the dashed vertical lines correspond to the linear frequencies (using the initial differential rotation profile) of the least damped mode for the respective azimuthal wavenumbers. The spatial structures of the linear and nonlinear eigenfunctions for cases (a) and (b) are shown in Fig.~\ref{fig:eig}.}
\label{fig:psd}
\end{figure*}

\subsection{Generation of higher harmonics from the fundamental $m=1$ unstable mode}

In fluid dynamics, a cascade describes the transfer of energy across different scales \citep[see, for instance,][]{Boning2023}. This transfer is controlled by the nonlinear advective term in the Navier-Stokes equation, $\left(\bm{u} \cdot \nabla\right)\bm{u}$, which couples different Fourier modes and redistributes energy between them. Decomposing the flow into azimuthal modes $\bm{u}(t,\theta,\phi) = \sum_m \bm{u}_m(t,\theta) \mathrm{e}^{\ii m \phi}$, the nonlinear term can be written as
\begin{equation}
    \left[ (\bm{u} \cdot \nabla) \bm{u} \right]_m  = \int_0^{2\pi} \sum_{m_1,m_2} \left[ \left( \bm{u}_{m_1} \textrm{e}^{\ii m_1 \phi} \right)  \cdot \nabla \right] \left(\bm{u}_{m_2} \textrm{e}^{\ii m_2 \phi} \right) \textrm{e}^{-\ii m \phi} \mathrm{d}\phi.
\end{equation}
We assume that the flow is dominated by a single unstable (fundamental) mode with wavenumber $m_0$ (in the present study $m_0=1$). Owing to the quadratic structure of the advective term, modes interact through the triadic condition $m = m_1 + m_2$. The self-interaction of the fundamental mode ($m_1 = m_2 = m_0$) generates the second harmonic $2m_0$, while its interaction with the complex conjugate ($m_1 = m_0$, $m_2 = -m_0$) modifies the axisymmetric component $m=0$. Subsequent interactions, such as $2m_0 + m_0 = 3m_0$, generate higher harmonics, leading to a cascade of energy \citep[see, for example,][their chapter~5]{Schmid2012}. 

\begin{figure*}[!htb]
\centering
  \begin{subfigure}{0.9\linewidth}
    \centering
    \includegraphics[width=\linewidth]{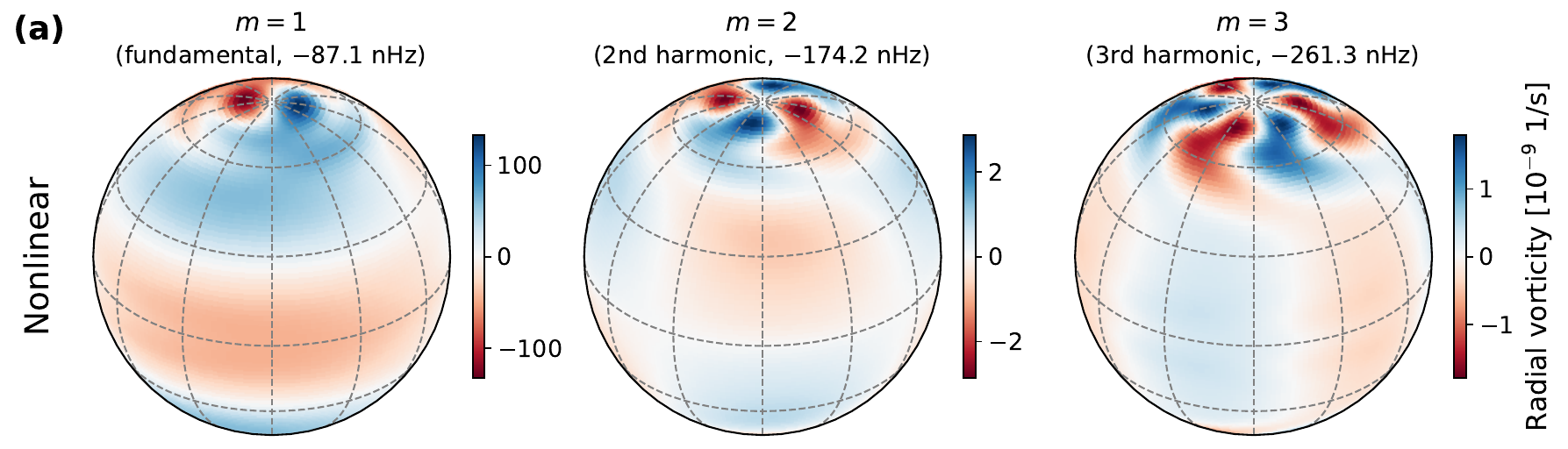}
  \end{subfigure} 
  \par\vspace{1\baselineskip}
  \begin{subfigure}{0.885\linewidth}
    \centering
    \hspace{-0.07\linewidth}
    \includegraphics[trim=0 0.05cm 0 0,clip,width=0.925\linewidth]{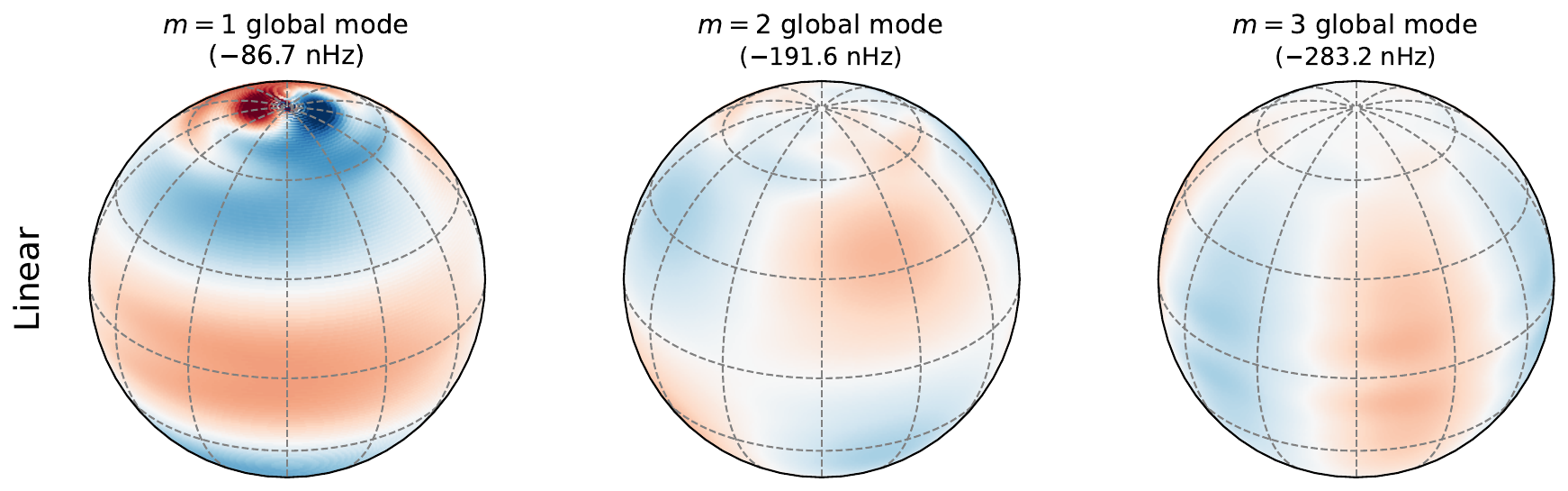}
  \end{subfigure}
\rule{0.9\linewidth}{0.5pt}
  \begin{subfigure}{0.9\linewidth}
    \centering
    \includegraphics[width=\linewidth]{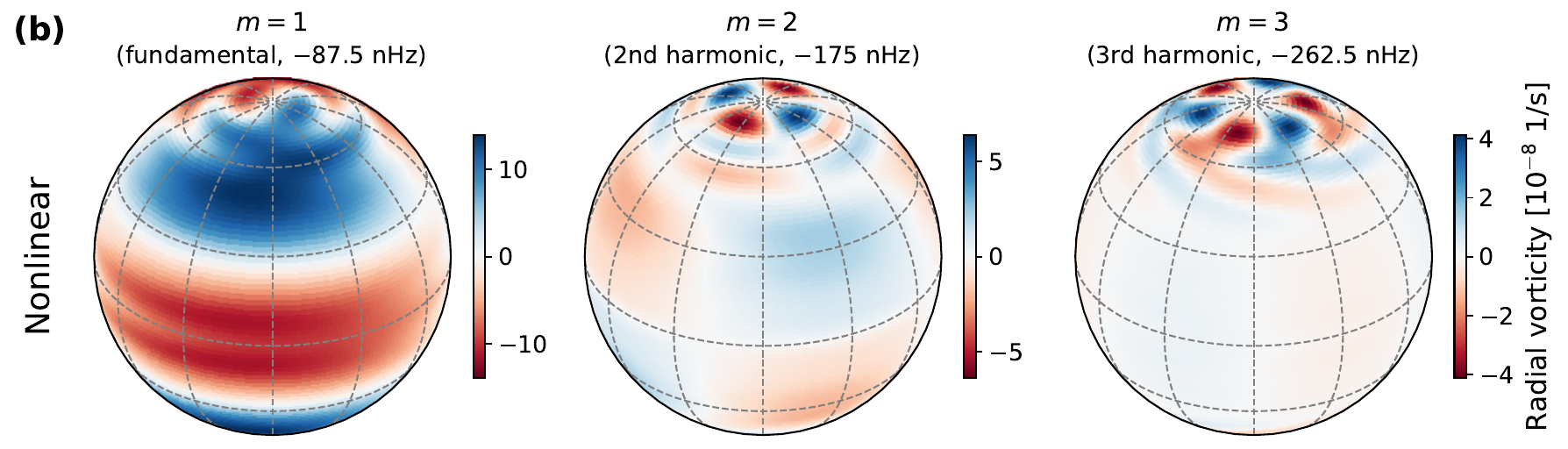}
  \end{subfigure}
  \par\vspace{1\baselineskip}
  \begin{subfigure}{0.885\linewidth}
    \centering
    \hspace{-0.07\linewidth}
    \includegraphics[trim=0 0.05cm 0 0,clip,width=0.925\linewidth]{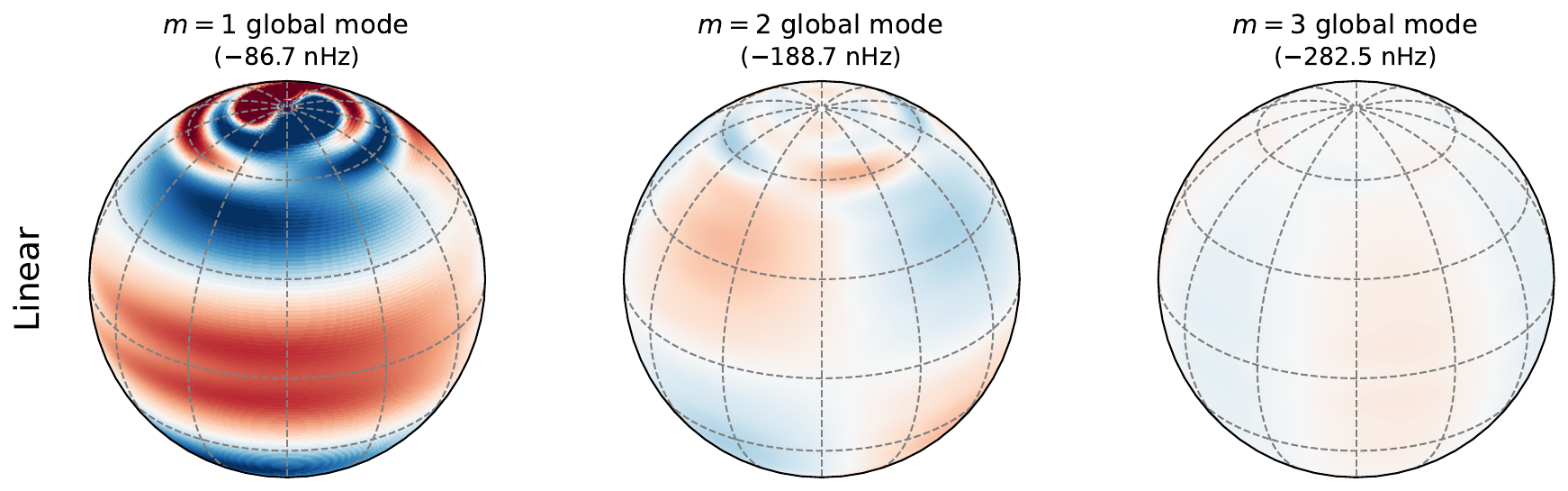}
  \end{subfigure}
  \caption{Radial vorticity eigenfunctions for $m=1,2,$ and 3 for (a) $E = 1.4 \times 10^{-3}$  and (b) $E = 4 \times 10^{-4}$. Top: Eigenfunctions from a DNS corresponding to the fundamental mode at (a) $\omega_{m = 1}/2\pi \approx -87.1$ nHz and (b) $-87.5$ nHz (in the Carrington reference frame), the second harmonic at $\omega_{m = 2} = 2 \omega_{m = 1}$, and the third harmonic at $\omega_{m = 3} = 3 \omega_{m = 1}$. These are extracted by filtering around the frequency of maximum power (see associated PSDs in Fig.~\ref{fig:psd}). Bottom: Least-damped linear global modes for $m = 1,2,$ and 3.} 
\label{fig:eig}
\end{figure*}

This cascade can be visualized by looking at the power spectral density (PSD) of the radial vorticity field defined as
\begin{equation}
    \mathrm{PSD}(\omega,m) := \langle|\tilde{Z}(\omega, \theta, m)|^2\rangle_\theta ,
\end{equation}
where $\langle\cdot\rangle_\theta$ denotes the average over the colatitude $\theta$, and $\tilde{Z}(\omega, \theta, m)$ is the two-dimensional discrete Fourier transform of $Z(t, \theta, \phi)$ in the saturation regime.
Figure~\ref{fig:psd} represents the PSD of the radial vorticity for the three Ekman numbers corresponding to Fig.~\ref{fig:snapshot_vorticity}. When only one mode is unstable ($E = 1.4 \times 10^{-3}$), the energy is concentrated in the $m=1$ mode at the frequency $\omega_{m=1}/2\pi \approx -87.1$ nHz, corresponding to nearly the eigenfrequency of the linear problem. This frequency shift can be estimated from the perturbative theories presented below (see Appendix~\ref{app:time_evolution}). There is then a cascade towards higher values of $m$ ($m=2$ at a frequency $2 \omega_{m=1}$, $m=3$ at $3 \omega_{m=1}$) with  a decrease in amplitude at each harmonic. 
We also extract the fundamental ($m=1$) state and the second and third harmonics ($m=2$ and $m=3$) from the saturated state  by filtering for a given $m$ in a narrow frequency band of width 30~nHz centered at $m \omega_{m=1}$ (see Fig.~\ref{fig:eig}a). The shape of the fundamental state is very close to the linear mode. However, the frequencies and spatial patterns of the harmonics differ significantly from the least damped $m=2$ and $m=3$ linear global modes. In particular, the harmonics exhibit with stronger power at high-latitudes compared to the linear eigenfunctions. 

For $E = 4 \times 10^{-4}$, even though multiple modes are unstable the $m=1$ mode still contains most of the energy (Fig.~\ref{fig:psd}b). However, its shape differs more from the linear mode than in the previous case (see Fig.~\ref{fig:eig}b). In this case,  the second and third harmonic have similar amplitudes.  
For even smaller viscosity ($E = 10^{-5}$), the situation is more complex with a competition between the $m=1$ and $m=2$ linearly-unstable modes, as can be seen in the movie associated with Fig.~\ref{fig:snapshot_vorticity}. The strongest $m=1$ mode in the saturation regime ($\omega/2\pi \approx -141.2$ nHz) is different from the one at the beginning of the simulation ($\omega/2\pi \approx -83.8$ nHz), and it corresponds to an eigenfrequency computed with the saturated rotation profile. At the end of the simulation, several modes have significant amplitudes, and this case can no longer be analyzed by simple linear theory or by the WNL theory presented in the next section.

\section{Weakly nonlinear (WNL) theories}\label{sec:weakly_nonlinear}

To gain some insight into the saturation mechanism of unstable modes and obtain their amplitudes without a recourse to nonlinear simulations, we propose using two classical perturbation methods: the multiscale \citep{Stuart1960} and amplitude expansion \citep{Watson1960} methods.  These asymptotic techniques incorporate selected nonlinear terms into a hierarchical framework of externally driven, linear equations. 
In the multiscale expansion, the small perturbation parameter $\varepsilon \ll 1$ is defined as the distance from the criticality.
In contrast, in the amplitude expansion, the time-dependent amplitude $A$ of the most unstable mode itself serves as the small parameter and must be sufficiently small so that truncation of the expansion is justified.
Although both approaches lead to the same amplitude equation --- the cubic Stuart–Landau equation, the computation and physical interpretation of the first Landau coefficient differ. In the multiscale expansion, this coefficient is obtained using the Fredholm solvability condition and represents the shift in growth rate and frequency of the critical eigenmode due to a slight detuning of the Ekman number. In contrast, in the amplitude expansion, the first Landau coefficient is simply the eigenvalue of the linear problem computed at a given Ekman number and hence represents the growth rate and frequency of the most unstable mode in the system.

These methods have been used extensively in hydrodynamic stability theory to assess the saturation of disturbances toward a finite-amplitude state and to determine the bifurcation behavior near criticality, and have been compared for example in \citep{Herbert1983, Fujimura1989, Zhang2016, Pham2018}. 
We present the applications of both approaches to our problem below (the general framework and details of the derivations are given in Appendix \ref{app:multiscale} and \ref{app:amplitude_expansion}).

\subsection{First method: Multiscale expansion}\label{sec:multiscale_exp}

In this expansion method, a fast time scale $\hat{t}$ (mode oscillations) and a slow time scale $\hat{T}$ are used. The expansion hierarchy is based on a small parameter $\varepsilon \ll 1$ which measures the departure from criticality, $E = \Ec + \varepsilon^2 E_2$  as well as the slow time scale $\hat{T}=\varepsilon^2 \hat{t}$. We expand the perturbation of the stream function as
\begin{equation}
     \hat{\Psi}(\hat{T},\hat{t}, \theta, \phi) = \varepsilon\,Q_1
    + \varepsilon^2\,Q_2 + \varepsilon^3\,Q_3+ \cdots,
\end{equation}   
where $Q_i = q_i + q_i^\ast$ so that $\hat{\Psi}$ is real. 
With this initialization, we follow the same procedural steps, as outlined in Appendix \ref{app:multiscale}.  At order $\mathcal{O}(\varepsilon)$, using the analogy to the general case \eqref{eq:linear_eig_}, we have
\begin{equation}
    q_1(\hat{T},\hat{t},\theta,\phi) = A(\hat{T}) \,\mathrm{e}^{-\ii (\hat{\omega} \hat{t} - m \phi)}\psi_{11}(\theta), 
\end{equation}
where $A(\hat{T})$ is a complex time-dependent amplitude, $\hat{\omega} = \omega/\Omega_\mathrm{ref} \in \mathbb{R}$ is the (dimensionless) frequency of the critical eigenmode $\psi_{11}(\theta)$ which satisfies
\begin{equation}
    \left(\hat{\omega} \Lm +\mathcal{L}_{m, \Ec}\right)\psi_{11}(\theta) = 0 \, ,
\label{eqn:leading_order_main}
\end{equation}
where  the linear operators $\Lm$ and $\mathcal{L}_{m, E}$ are defined as
\begin{align}
    \Lm &:= \frac{1}{\sin\theta}\frac{\partial}{\partial \theta}\left(\sin\theta \frac{\partial}{\partial \theta}\right) - \frac{m^2}{\sin^2\theta} \, , \label{eqn:L_m} \\[4pt]
    \mathcal{L}_{m, E} &:= -( m\delta + 2 \ii E)\Lm + \frac{m}{\sin\theta}\frac{\partial \hat{Z}_0}{\partial\theta} - \ii E \Lm^2 \, .
\label{eqn:L_A}
\end{align}

At order ${\cal{O}}(\varepsilon^2)$ we encounter nonlinear interactions of the perturbations that modify the base flow and generate higher-harmonic disturbances. The quadratic nonlinear term $\mathcal{N}[Q_1, Q_1]$ (see Eq.~\eqref{eq:non-linear_part_pde}) creates harmonics at $(\omega=0,m=0)$, $(2\omega,2m)$ and $(-2\omega,-2m)$ and we seek for a solution of the form
\begin{equation}
    q_2(\hat{T},\hat{t},\theta,\phi) = |A(\hat{T})|^2 \psi_{20}(\theta) + A^2(\hat{T}) \psi_{22}(\theta) \mathrm{e}^{-2 \ii(\hat{\omega} \hat{t} -m \phi)}.
\label{eqn:q2}
\end{equation}

At order ${\cal{O}}(\varepsilon^3)$, due to the nonlinear interactions $\mathcal{N}[Q_1, Q_2]$ and $\mathcal{N}[Q_2,Q_1]$, we obtain a correction to the fundamental mode and a third harmonic and the solution $q_3$ takes the form
\begin{align}
    q_3(\hat{T},\hat{t}, \theta, \phi) &= |A(\hat{T})|^2A(\hat{T}) \psi_{31}(\theta)\mathrm{e}^{-\ii(\hat{\omega}\hat{t} - m \phi)}  \nonumber \\ 
    &\hspace{2cm}+ A^3(\hat{T}) \psi_{33}(\theta) \mathrm{e}^{-3\ii(\hat{\omega} \hat{t} -m \phi)}.
\label{eqn:q3}
\end{align}
The horizontal components $\psi_{kj}$ (for $\psi_{20}, \psi_{22}, \psi_{33}$) are obtained by using the expressions for $q_2$ and $q_3$ into Eq.~\eqref{eqn:psi_eqn} leading to
\begin{equation}
    \left( j\hat{\omega} \mathbb{L}_{jm} +  \mathcal{L}_{jm, \Ec} \right) \psi_{kj}(\theta) = \ii f_{kj}(\theta), \label{eq:multi-scale-gen}
\end{equation}
where the right-hand sides $f_{kj}$ are given in Appendix \ref{app:multiscale}.
For $\psi_{31}$, the slow time-scale enters the asymptotic expansion 
via $\partial_{\hat{t}}\to \partial_{\hat{t}} + \varepsilon^2 \partial_{\hat{T}},$ and yields
\begin{align}
    |A|^2A \left(\hat{\omega}\Lm + \mathcal{L}_{m,\Ec} \right)\psi_{31}(\theta) &= \ii |A|^2A f_{31}(\theta) - \ii \frac{\partial A}{\partial \hat{T}} \Lm \psi_{11}(\theta) \nonumber \\ 
    &\hspace{0.25cm} + \ii A E_2\left(\Lm + 2\right) \Lm \psi_{11}(\theta) . \label{eqn:3rd_main}
\end{align}
The left-hand side corresponds to our eigenvalue problem and is thus singular. According to the Fredholm alternative theorem, this equation is solvable if  the forcing term on the right-hand side is orthogonal to the null space of the adjoint operator.
Taking the inner product of Eq. (\ref{eqn:3rd_main}) with the adjoint eigenfunction $\psi_{11}^\dagger$ yields the celebrated cubic \textit{Stuart-Landau equation} \citep[see, for instance,][]{Landau1944, Stuart1960}  
\begin{equation}
    \frac{\partial A}{\partial \hat{T}} = -\ii \alpha A(\hat{T}) - \ii\Gamma |A(\hat{T})|^2A(\hat{T}) ,
\label{eqn:Stuart_Landau_eq}
\end{equation}
where $\alpha$ and $\Gamma$ are refereed to as the Landau coefficients which are defined as
\begin{align}
    \alpha := E_2  \xi := E_2\frac{\langle \ii(\Lm +2)\Lm\psi_{11}, \psi_{11}^\dagger\rangle}{\langle\Lm\psi_{11}, \psi_{11}^\dagger\rangle} , \quad
    \Gamma := \frac{\langle \ii f_{31}, \psi_{11}^\dagger\rangle}{\langle\Lm\psi_{11}, \psi_{11}^\dagger\rangle} .
\label{eqn:sigma_beta_main}
\end{align}
Once the second Landau coefficient is known, the correction to the fundamental can be obtained by solving Eq.~\eqref{eqn:3rd2a}.

\subsection{Second method: Amplitude expansion}\label{sec:amplitude_exp}

In the amplitude expansion, we use the Ansatz that the normalized perturbation of the stream function $\hat{\Psi}$ can be written as
\begin{equation}
    \hat{\Psi}(\theta,\phi, \hat{t}) =  A(\hat{t}) \psi_{11}(\theta) \mathrm{e}^{\ii m\phi} + \textrm{c.c.} , 
\label{eqn:psi11}
\end{equation}
where c.c. denotes the complex conjugate. 
Substituting this expansion into Eq. \eqref{eqn:psi_eqn}, we obtain
\begin{align}
    &\left(\mathbb{L}_m \psi_{11}(\theta) \frac{\dd A}{\dd \hat{t}} - \ii \mathcal{L}_{m, E}\psi_{11}(\theta) A (\hat{t}) \right) \mathrm{e}^{\ii m \phi} + \textrm{c.c.} \nonumber\\
   &\hspace{2cm} = |A(\hat{t})|^2 f_{20}(\theta) + \left(A(\hat{t})^2 f_{22}(\theta) \mathrm{e}^{2 \ii m \phi} + \textrm{c.c.}\right) , 
\label{eqn:amp1b_main}
\end{align}
where the linear operators $\mathbb{L}_m$ and $\mathcal{L}_{m, E}$ are defined in Eqs. (\ref{eqn:L_m}) and (\ref{eqn:L_A}) and the expressions for nonlinear functions $f_{20}(\theta)$ and $f_{22}(\theta)$ are given in Eqs. (\ref{eqn:f20}) and (\ref{eqn:f22}) with $\psi_{11}$ computed at the Ekman number $E$ (not at $\Ec$ as in the multiscale expansion).

Keeping only the linear terms in amplitude $A(\hat{t})$ in Eq. (\ref{eqn:amp1b_main}) and using the wave ansatz $A(\hat{t}) = A_0 \mathrm{e}^{-\ii \sigma \hat{t}/\Omega_\mathrm{ref}}$, with $\sigma$ complex and $A_0 = A(\hat{t} = 0)$, we get the leading-order equation
\begin{equation}
    \left(\frac{\sigma}{\Omega_\mathrm{ref}}\mathbb{L}_m + \mathcal{L}_{m, E}\right) \psi_{11}(\theta) = 0,
\label{eqn:lin_prob_main}
\end{equation}
which corresponds to the linear eigenvalue problem and gives access to the most unstable eigenmode with complex eigenvalue $\sigma$ and eigenfunction $\psi_{11}$.

Similarly to the terms in $\varepsilon^2$ and $\varepsilon^3$ in the multiscale expansion, we collect the terms of order two and three in amplitude and write 
\begin{align}
    \hat{\Psi}(\theta,\phi, \hat{t}) = & A(\hat{t}) \psi_{11}(\theta) \mathrm{e}^{\ii m\phi}  +  |A(\hat{t})|^2 \psi_{20}(\theta) + A^2(\hat{t}) \psi_{22}(\theta) \mathrm{e}^{2 \ii m \phi} \nonumber \\
    &  + |A(\hat{t})|^2 A(\hat{t}) \psi_{31}(\theta) \mathrm{e}^{\ii m\phi} + A^3(\hat{t}) \psi_{33}(\theta) \mathrm{e}^{3 \ii m \phi} + \textrm{c.c.} .
\end{align}
The terms of order two (resp. three) are necessary to compensate the nonlinear interactions of the fundamental with itself (resp. with the second harmonic). As in the multiscale expansion, the base flow is modified and second and third harmonics are created due to the nonlinear interactions. The spatial distribution $\psi_{kj}$ is obtained by solving
\begin{equation}
   \left(\frac{j \sigma_R + \ii k \sigma_I}{\Omega_\mathrm{ref}}\mathbb{L}_{jm} + \mathcal{L}_{jm,E}\right)\psi_{kj} = \, i f_{kj}, 
\label{eqn:higher_order_amp_exp}
\end{equation}
where $\sigma = \sigma_R + \ii \sigma_I$ and the explicit expressions of the right-hand sides can be found in Appendix~\ref{app:multiscale}. In this case,  the linear operators on the left-hand sides are non-singular, and thus invertible. In order to truncate the expansion in amplitude, we need to assume the time evolution of the disturbance amplitude can not be determined from linear theory alone, instead we need to add one more term in the evolution equation as follows \citep[see, for example,][]{Crouch1993, Pham2018}
\begin{equation}
    \frac{\mathrm{d} A(\hat{t})}{\mathrm{d} \hat{t}} = -\ii \left( \frac{\sigma}{\Omega_\mathrm{ref}} A(\hat{t}) + \frac{\beta}{\Omega_\mathrm{ref}} |A(\hat{t})|^2 A(\hat{t}) \right) .
\label{eqn:cubic_SL_eqn1_main}
\end{equation}
This is the Stuart-Landau equation \citep[see, for example,][]{Landau1944, Stuart1960}, where $\sigma$, the first Landau coefficient, is the linear complex eigenvalue and $\beta$ is the second Landau constant which is yet to be determined. 
Using Eq.~\eqref{eqn:cubic_SL_eqn1_main} instead of the wave ansatz affects only the equation for $\psi_{31}(\theta)$ which becomes
\begin{equation}
    \left(\frac{\sigma+2i\sigma_I}{\Omega_\mathrm{ref}}\mathbb{L}_m + \mathcal{L}_{m, E}\right) \psi_{31}(\theta) = \ii f_{31}(\theta) - \frac{\beta}{\Omega_\mathrm{ref}} \mathbb{L}_m \psi_{11}(\theta) .
\label{eqn:f31_landau_main}
\end{equation}
In the resonant case ($\sigma_I = 0$), the second Landau constant is uniquely defined and can be obtained by using the Fredholm-alternative theorem. However, when $\sigma_I \neq 0$,  the value of the Landau constant is not uniquely determined since the inhomogeneous system (\ref{eqn:f31_landau_main}) is solvable for any value of $\beta$ and an additional constrain has to be added. \citet{Crouch1993} propose to assume weighted orthogonality between $\psi_{11}$ and $\psi_{31}$, that is,
\begin{equation}
    \langle \psi_{11} , \, \psi_{31}\rangle_\mathcal{M} := \psi_{11}^H\ \mathcal{M} \psi_{31} = 0 ,
\label{eqn:extra_cond}
\end{equation}
where $\mathcal{M}$ is a positive definite matrix and $(\cdot)^H$ denotes the Hermitian transpose.  \citet{Pham2018} showed that this condition is equivalent to the solvability condition when $\sigma_I \to 0$.  Several choices are possible for $\mathcal{M}$ \citep[][their section~4]{Pham2018}. Here, we choose the simplest option: $\mathcal{M} = I$, the identity matrix. Then the Landau constant $\beta$ and the function $\psi_{31}$ are uniquely determined by solving the extended system
\begin{align}
    &\tilde{\mathcal{L}} \tilde{\psi}_{31} = \ii \tilde{f}_{31} , \quad     \tilde{\psi}_{31} = \left[ \begin{array}{c} \psi_{31} \\ \beta/\Omega_\mathrm{ref} \end{array} \right] , \quad 
    \tilde{f}_{31} = \left[ \begin{array}{c}  f_{31}  \\ 0 \end{array} \right]  ,  \\
   &\textrm{where} \quad \tilde{\mathcal{L}} = \left[
    \begin{array}{cc}
    (\sigma+2i\sigma_I)/\Omega_\mathrm{ref} \mathbb{L}_m +\mathcal{L}_{m, E} &  \mathbb{L}_m \psi_{11} \\
    \psi_{11}^H \mathcal{M} & 0
    \end{array}
    \right] .
    \label{eqn:augmented_sys_main}
\end{align}

Although the multiscale and amplitude expansion methods rely on different perturbation frameworks, both lead to the same Stuart--Landau normal form for the mode amplitude. The apparent difference between the corresponding Landau coefficients arises from the use of different small parameters and normalizations. The two sets of coefficients are related by a simple rescaling. In the limit $E\to \Ec$, one finds
\begin{equation}
\frac{\sigma}{\Omega_{\rm ref}} \to \alpha, \quad \frac{\beta}{\Omega_{\rm ref}} \to \Gamma,
\end{equation}
so that the two approaches become asymptotically equivalent. This is confirmed by the close agreement between the two methods near onset (Figs.~\ref{fig:amplitude1} and~\ref{fig:ratio_amp}).

\subsection{Symmetries}

The symmetries of the modes play an important role in their identification in observations \citep{Gizon2021}. Here we show that the symmetries of the higher harmonics are entirely determined by the symmetry of the unstable fundamental mode.

The functions $f_{20}$ and $f_{22}$ defined by Eqs. (\ref{eqn:f20}) and (\ref{eqn:f22}) are bilinear combinations of $\psi_{11}$ and $\zeta_{11}$, including their latitudinal derivatives (which has the opposite symmetry) and complex conjugates. They are thus always north-south (NS) antisymmetric implying that the functions $\psi_{20}$ and $\psi_{22}$ are always NS antisymmetric.

The functions $f_{31}$ and $f_{33}$ include the terms which are product of functions with same NS symmetry implying that the correction to the fundamental and the third harmonic are NS symmetric (same symmetry as the fundamental mode). In most of our simulations, the most unstable mode is the high-latitude $m=1$ which is NS symmetric and so is the third harmonic.

\subsection{Numerical implementation}
The different ODEs obtained in the multiscale and amplitude expansions (see Subsections \ref{sec:multiscale_exp} and \ref{sec:amplitude_exp}) are discretized by projecting $\psi_{ij}$ into the associated Legendre polynomials $P_\ell^j$. The first step is to solve the leading-order equation in both the expansion methods (Eq. (\ref{eqn:leading_order_main}) in multiscale and (\ref{eqn:lin_prob_main}) in amplitude expansion), where in multiscale expansion, it is solved exactly at the critical Ekman number and in the amplitude expansion, it is solved at the Ekman number considered. From $\psi_{11}$ it is possible to compute the right-hand sides $f_{20}$ and $f_{22}$ for second-order equations ((\ref{eqn:2nd1}) and (\ref{eqn:2nd2}) in multiscale and (\ref{eqn:2nd_order1a}) and (\ref{eqn:2nd_order2a}) in amplitude expansion) and to solve the linear systems to obtain $\psi_{20}$ and $\psi_{22}$. The linear operators $\mathbb{L}_m$ and $\mathcal{L}_{m,E}$ for different values of $m$ and $\sigma$ are discretized as in \citet{Fournier2022}. From $\psi_{11}$, $\psi_{20}$ and $\psi_{22}$ it is possible to compute the right-hand sides $f_{31}$ and $f_{33}$ for third-order equations ((\ref{eqn:3rd1}) and (\ref{eqn:3rd2}) in multiscale and (\ref{eqn:3rd_order1a}) and (\ref{eqn:3rd_order2a}) in amplitude expansion). In multiscale expansion method, we can then compute the two Landau coefficients using Eq. (\ref{eqn:sigma_beta_main}) and finally obtain from Eq. (\ref{eq:multi-scale-gen}) the spatial distribution of the third harmonic $\psi_{33}$. On the other hand in amplitude expansion, the first Landau coefficient is just the linear growth rate of the most unstable mode at the considered Ekman number and we have to solve the augmented system (\ref{eqn:augmented_sys_main}) to obtain the second Landau coefficient $\beta$ and the third harmonic $\psi_{33}$. Hence for both the expansion methods, the computational cost is only the resolution of four linear systems and the linear eigenvalue problem. This should be done only at the critical Ekman number for the multiscale expansion and repeated at each Ekman for the amplitude expansion.

\section{Comparison of WNL theories with DNS}
\subsection{Extraction of Landau coefficients from DNS}
\begin{figure}
    \centering
    \includegraphics[width=0.9\linewidth]{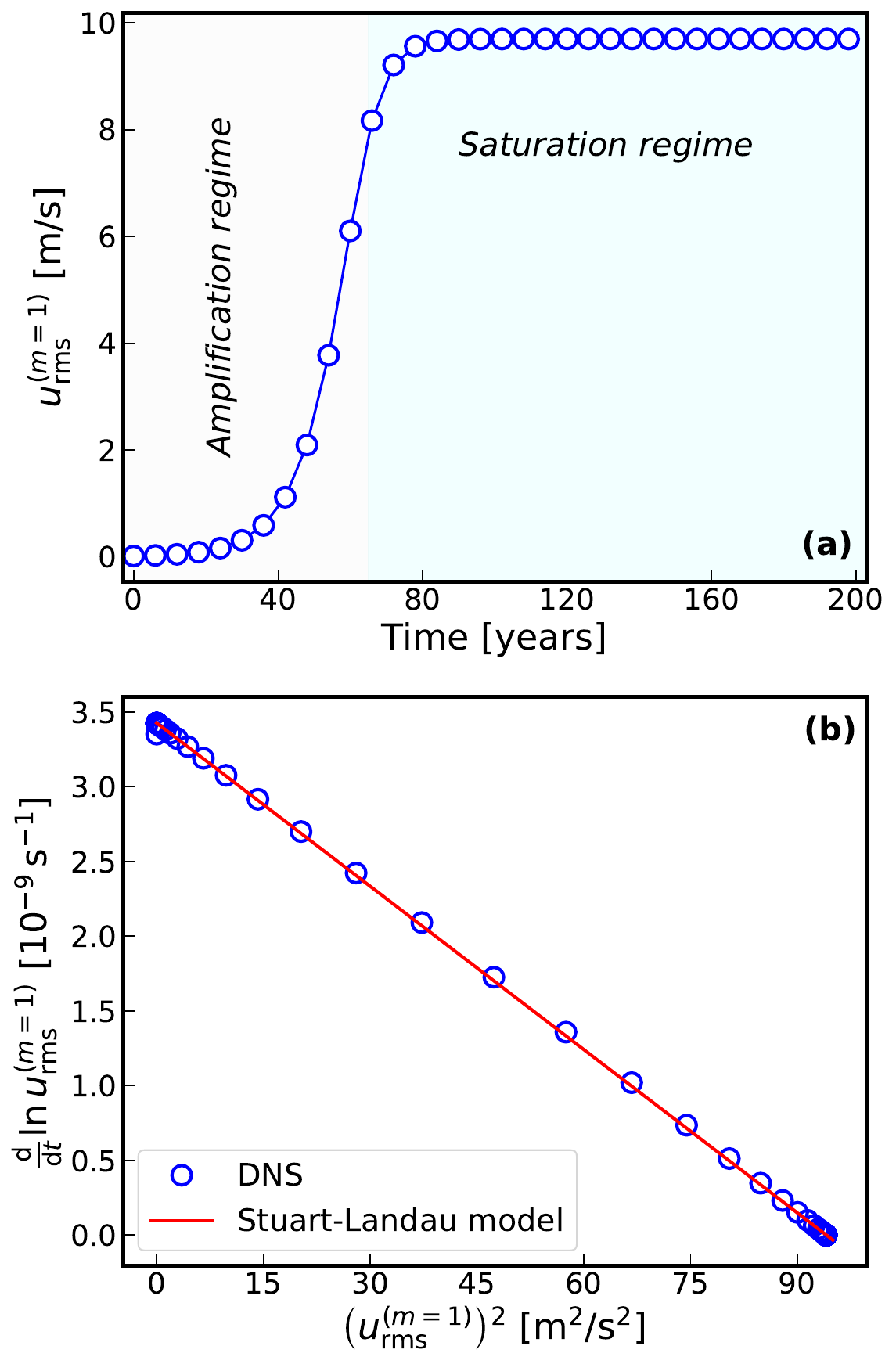}
\caption{RMS velocity of the $m = 1$ mode for $E = 1.4 \times 10^{-3}$; (a) Time evolution of the RMS velocity demonstrating an exponential growth (amplification regime) followed by nonlinear saturation (saturation regime). (b) Time derivative of $\ln u^{(m  =1)}_\mathrm{rms}$ versus $\left(u^{(m = 1)}_\mathrm{rms}\right)^2$ and linear fit according to Eq.~\eqref{eqn:lin_func_uavg}.}
\label{fig:DNS}
\end{figure}

We use the $m = 1$ filtered RMS velocity $u^{(m = 1)}_\mathrm{rms}$ as our Landau variable. The Stuart-Landau equation can be written as
\begin{equation}
    \frac{\mathrm{d}}{\mathrm{d}T}\ln u^{(m = 1)}_\mathrm{rms}(T) = \tilde{\sigma}_I + \tilde{\beta}_I  \, \left(u^{(m = 1)}_\mathrm{rms}(T)\right)^2 , 
\label{eqn:lin_func_uavg}
\end{equation}
where the Landau coefficients $\tilde{\sigma}_I$ and $\tilde{\beta}_I$ have units of frequency ($1/\mathrm{s}$) and inverse of kinematic viscosity ($\mathrm{s}/\mathrm{m}^2$), respectively, and need to be related to $\alpha_I = \Im(\alpha)$ and $\Gamma_I = \Im(\Gamma)$ from Eq.~(\ref{eqn:sigma_beta_main}).  
Using the velocity-stream function relation (\ref{eqn:straamfun}), the relation between $u^{(m = 1)}_\mathrm{rms}(T)$ and $|A(T)|$  is
\begin{equation}
    u^{(m = 1)}_\mathrm{rms}(T) = \overline{U}_1 \,  \, |A(T)| ,
\label{eqn:urmsA}
\end{equation}
where $\overline{U}_1$ is the norm of the  flow eigenfunction for $m=1$ (see also Eq.~\eqref{eqn:straamfun1} for a definition of the total flow)
\begin{equation}
     \overline{U}_1^2  = r^2 \Omega_{\rm ref}^2 \int_0^\pi \left(\frac{1}{\sin^2\theta}|\psi_{11}|^2 
    + \left|\frac{\dd \psi_{11}}{\dd \theta}\right|^2\right) \sin\theta \mathrm{d}\theta .
\label{eqn:integral}
\end{equation}
Combining Eqs. (\ref{eqn:urmsA}) in (\ref{eqn:lin_func_uavg}) yields a relationship between the Landau coefficients which is given as 
\begin{equation}
    \frac{\tilde \sigma_I}{\Omega_{\rm ref}} = \alpha_I \, , \quad \frac{\tilde \beta_I}{\Omega_{\rm ref}} = \frac{\Gamma_I}{\overline{U}_1^2} .
\label{eqn:Landau_coeff_reln}
\end{equation}
It is thus possible to obtain the Landau coefficients $\tilde{\sigma}_I$ and $\tilde{\beta}_I$ directly from the DNS by performing a linear fit between  the time derivative of $\ln{u^{(m = 1)}_\mathrm{rms}}$ and $\left(u^{(m = 1)}_\mathrm{rms}\right)^2$ as shown in Fig.~\ref{fig:DNS} for an Ekman number of $E = 1.4 \times 10^{-3}$. The DNS data are well represented by this linear fit, and the slope is negative confirming the supercritical nature of the bifurcation. The amplitude variations with time showing the amplification and saturation regime can be obtained from the (exact) solution of the Stuart-Landau equation (for details, see Appendix~\ref{app:time_evolution}).

\subsection{Ekman number dependence of RMS velocity amplitudes}

\begin{figure}[t]
    \centering
    \includegraphics[width=0.92\linewidth]{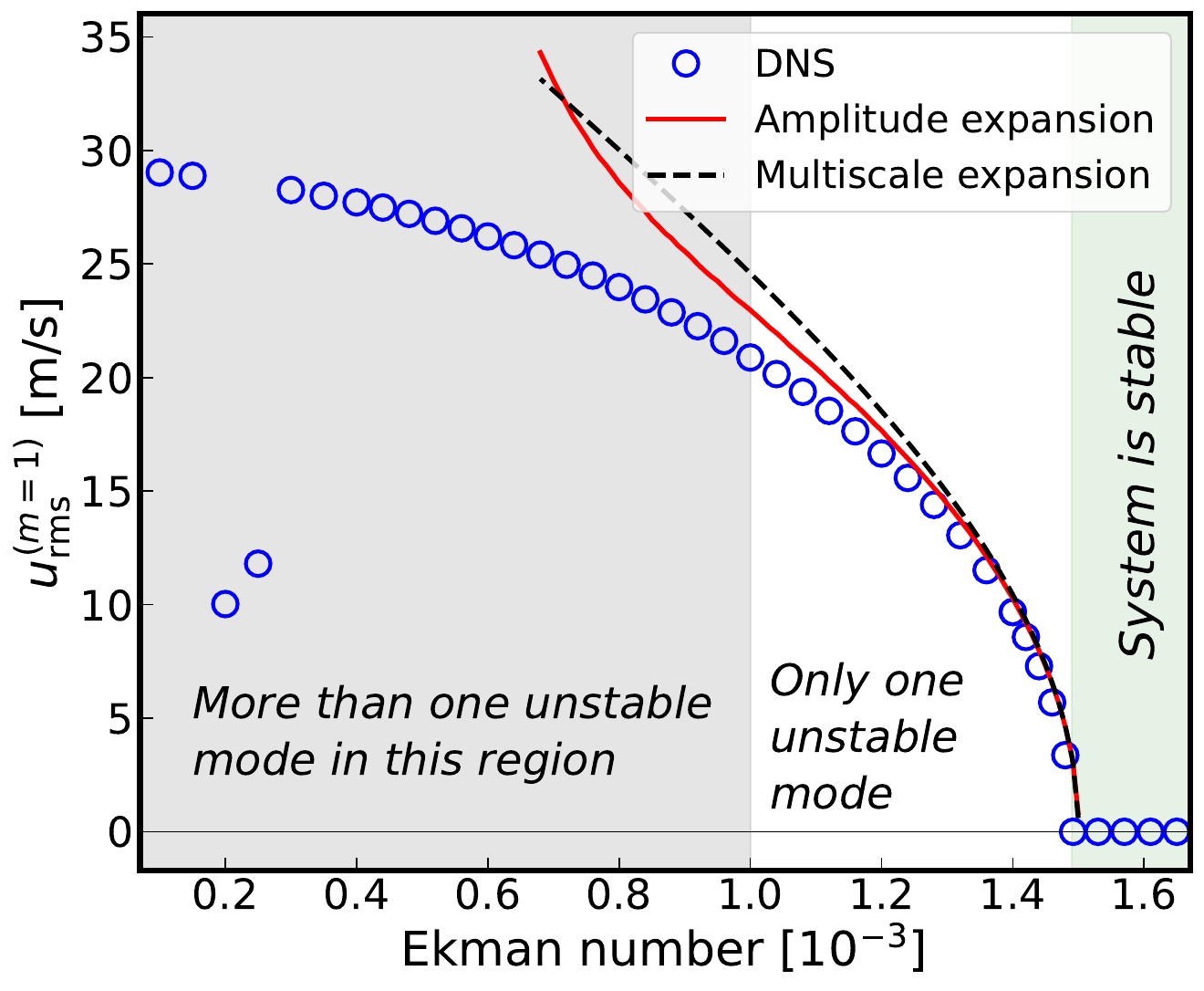}
\caption{Ekman number dependence of the saturated RMS velocity amplitudes for $m = 1$, as obtained from DNS and WNL theory. The gray shaded region indicates where multiple modes become unstable and the simple WNL theory presented here is not valid. The $m=1$ mode is dominant in the saturation regime except for  $E = 2.5 \times 10^{-4}$ and $2 \times 10^{-4}$. In these two cases, the $m=2$ mode is dominant which explains the low RMS velocity amplitude of the $m=1$ mode.} 
\label{fig:amplitude1}
\end{figure} 

We use the procedure described above to obtain the Landau coefficients as a function of the Ekman number  (Table \ref{table:Ekman_sigmaI_betaI}). For all the considered Ekman numbers, we obtain positive first and  negative second Landau coefficients which indicates the supercritical nature of the system. 

\begin{figure}[t]
    \centering
    \includegraphics[width=\linewidth]{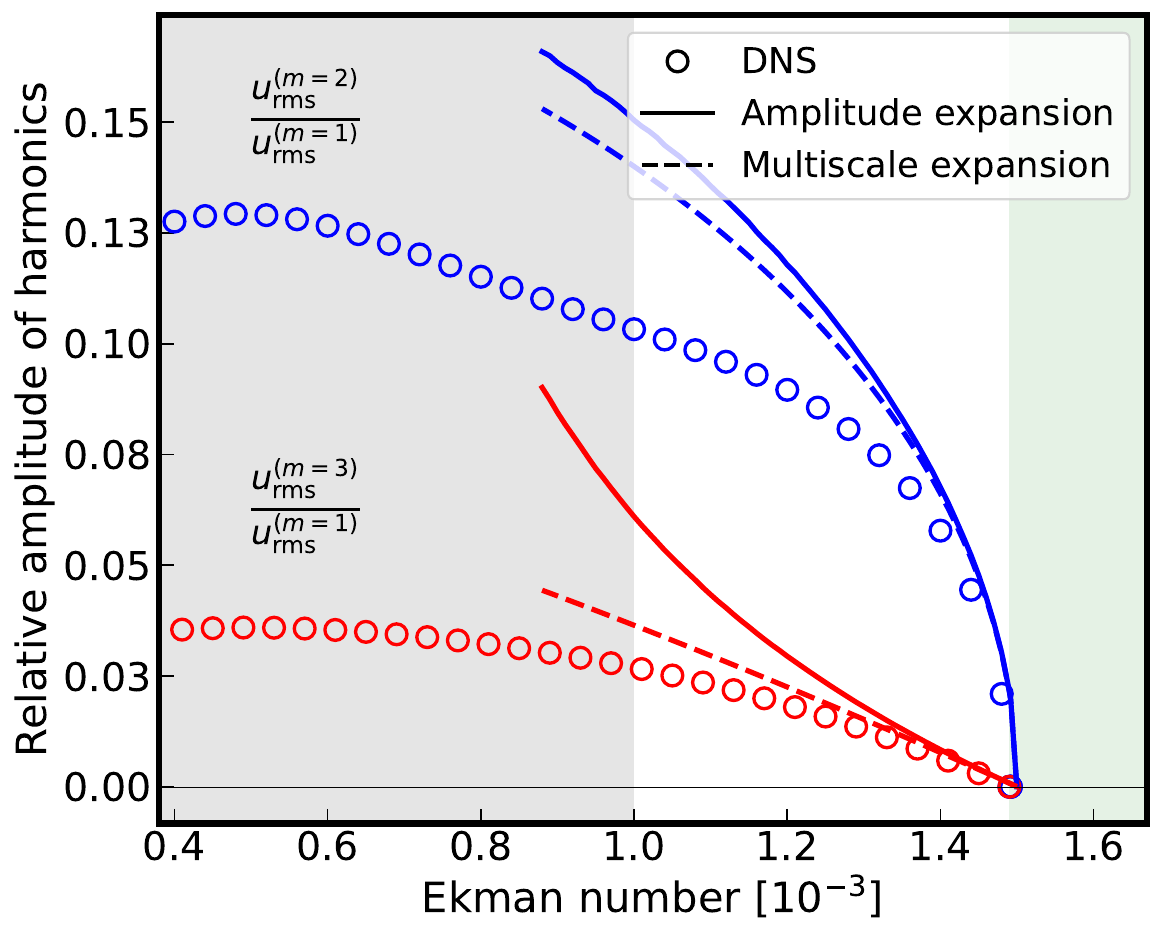}
\caption{Relative amplitude of the second and third harmonics, $u_\mathrm{rms}^{(m=2)}/u_\mathrm{rms}^{(m=1)}$ and $u_\mathrm{rms}^{(m=3)}/u_\mathrm{rms}^{(m=1)}$. Empty dots represent the DNS results, solid and dashed curves represent the (amplitude and multiscale expansion respectively) WNL results. The gray shaded region indicates the parameter range where WNL theory is not expected to provide reliable results due to the presence of multiple unstable modes. The system remains stable in the green-shaded region.}
\label{fig:ratio_amp}
\end{figure}

Since the amplitudes of the higher harmonics scale as $|A|^m$, we can relate the RMS velocity of the flow field of azimuthal order $m$ to the amplitude $|A|$. In particular, for the multiscale expansion we obtain
\begin{equation}
    u^{(m)}_\mathrm{rms} = \overline{U}_m |A|^m = \overline{U}_m \left(-\frac{\alpha_I}{\Gamma_I}\right)^{m/2} = \overline{U}_m \left(-\frac{\xi_I}{\Gamma_I}\right)^{m/2} (\Ec - E)^{m/2},
\label{eqn:urms_scaling}
\end{equation}
where $\overline{U}_m$ is defined similarly to Eq.~\eqref{eqn:integral} by replacing $\psi_{11}$ with the stream function for azimuthal order $m$, $\psi_{mm}$.  
Figure \ref{fig:amplitude1} shows the Ekman number dependence of the RMS velocity amplitude for $m = 1$ using DNS and the WNL theories. The optimal fit $\gamma (\Ec - E)^{1/2}$ to the DNS data yields, $\gamma^{\rm DNS} \approx 996$ m/s while $\gamma^{\rm WNL} \approx 1104.6$ m/s for the multiscale expansion. Equation~(\ref{eqn:urms_scaling}) corresponds to the relation obtained by \citet{Bagheri2013} where they assumed that the norm of  $\overline{u}_{\rm rms}^{(m)}$ is of order 1. In our case, we obtain 
\begin{equation}
    \frac{u^{(m=2)}_\mathrm{rms}}{u^{(m = 1)}_\mathrm{rms}} = \mathcal{C}_2 ( \Ec - E)^{1/2} \quad \textrm{and} \quad \frac{u^{(m=3)}_\mathrm{rms}}{u^{(m = 1)}_\mathrm{rms}} = \mathcal{C}_3 (\Ec - E),
\label{eqn:urms_ratio_scaling}
\end{equation}
with $\mathcal{C}^\mathrm{WNL}_2 \approx 6.1$ and $\mathcal{C}^\mathrm{WNL}_3 \approx 69.9$. Assuming this scaling and fitting to the DNS data, we obtained $\mathcal{C}_2^{\rm DNS} \approx 5.8$ and $\mathcal{C}_3^{\rm DNS} \approx 69.1$. A representation of these approximations of the ratio, directly obtained from the DNS, is shown in Fig.~\ref{fig:ratio_amp}. Close to the critical Ekman, the scaling with the Ekman is well captured, and the WNL theories provide a good approximation of the amplitude ratio. The quality of the approximation deteriorates with the distance to $\Ec$, but remains within 35\% for both $m=2$ and $m=3$ of the DNS value even for $E=10^{-3}$.  

\subsection{Spatial structure of fundamental and harmonics}

\begin{figure}
    \centering
\includegraphics[width=0.85\linewidth]{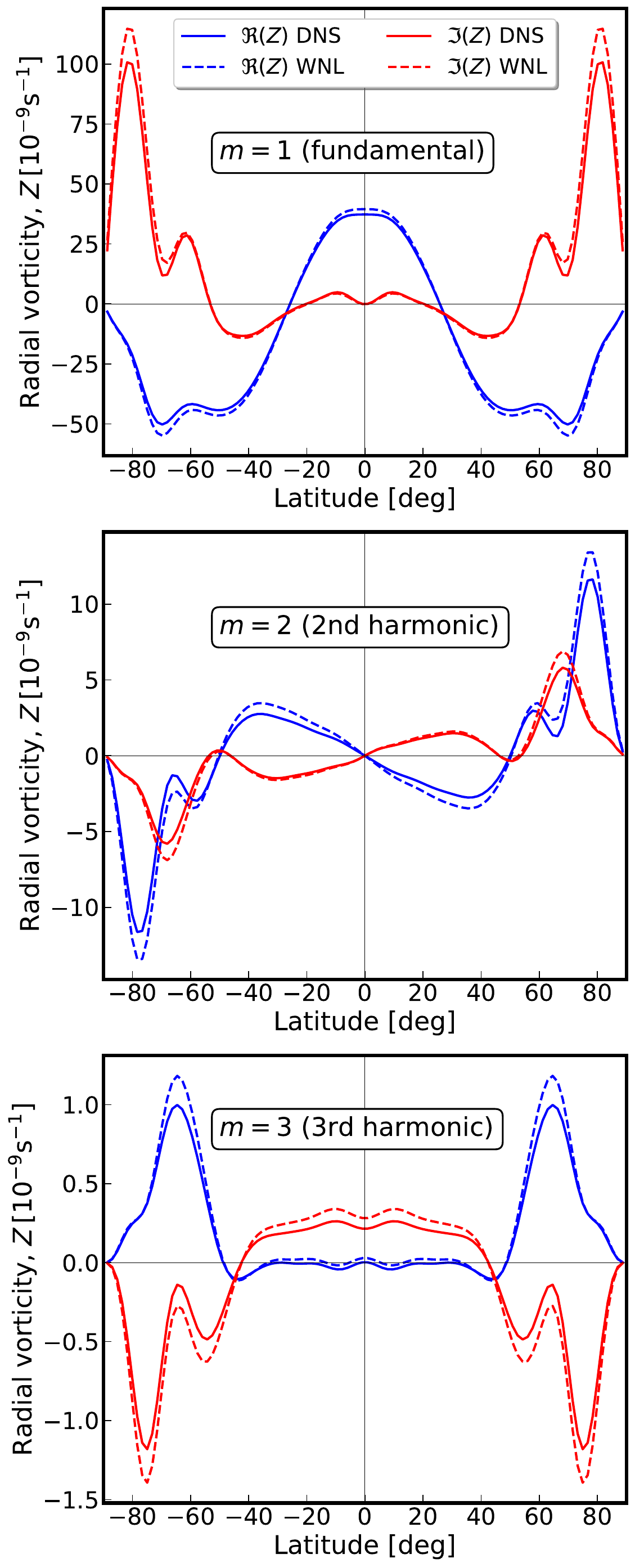}
\caption{Radial vorticity $Z$ for $m = 1, 2$ and $3$ from the DNS (solid curves) and the WNL theory (dashed curves) for an Ekman number $E = 1.4 \times 10^{-3}$.}
\label{fig:vorticity}
\end{figure}

Once the amplitude is determined, it is possible to obtain the eigenfunctions and their harmonics from the perturbation methods. Figure~\ref{fig:vorticity} compares the radial vorticity, $Z$ for $m = 1, 2$ and $3$ from the multiscale expansion and the DNS for an Ekman number $E = 1.4 \times 10^{-3}$ (the amplitude expansion gives a similar profile). For $m = 1$, we have adjusted the phase such that the imaginary part of the eigenfunction vanishes at the equator. The shape of the fundamental and its first two harmonics are well reproduced by WNL theory. However, the differences with the DNS increases with the distance to criticality. As the amplitudes of the harmonics scale as $A^m$, the discrepancy between WNL theory and DNS increases for the harmonics compared to the fundamental. 

\subsection{WNL estimates of Reynolds stress and equilibrium  differential rotation} \label{sect:WNL_rot}

\begin{figure}
    \centering
    \includegraphics[width=0.9\linewidth]{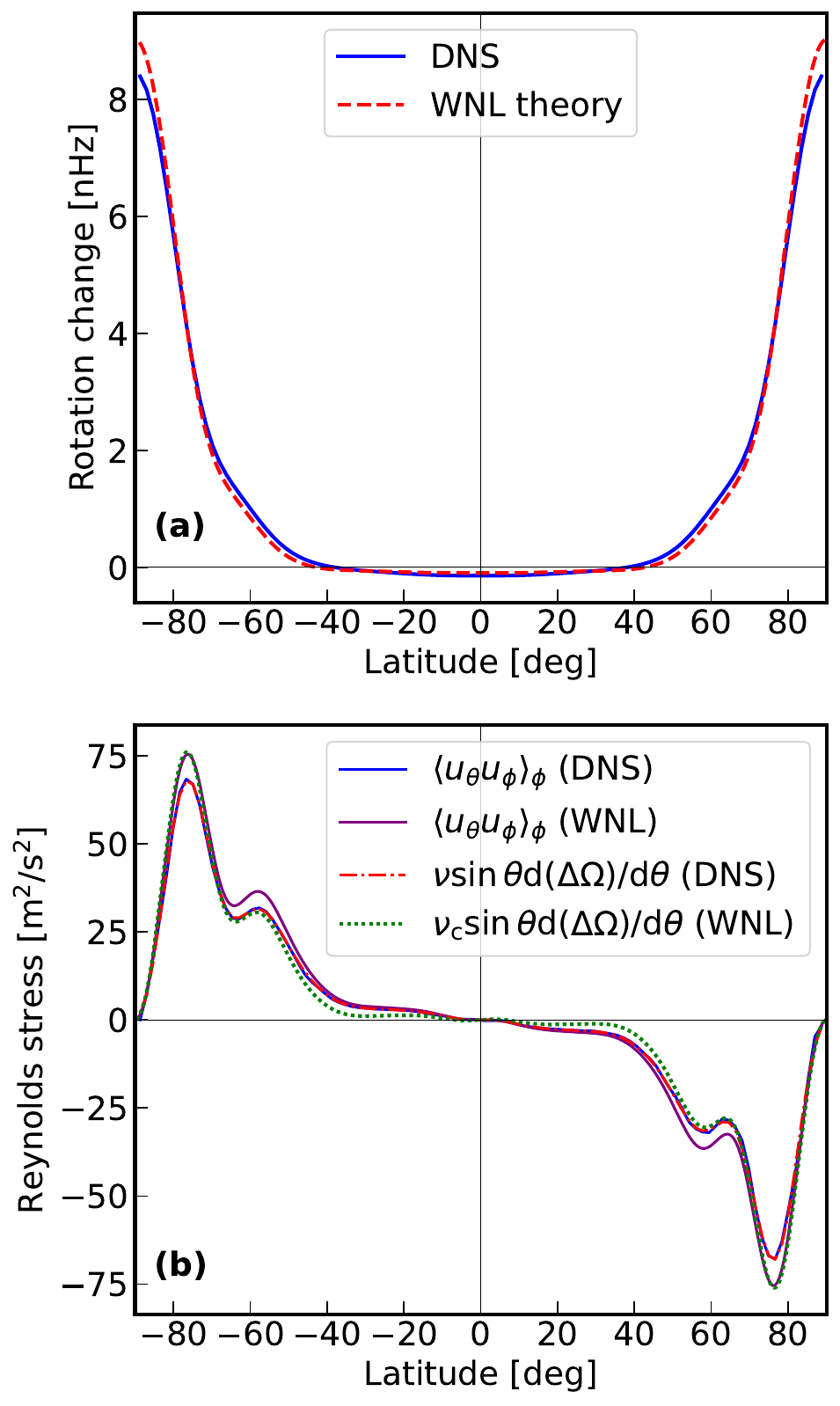}
\caption{(a) Change in differential rotation at equilibrium, $\Delta\Omega=\Omega(\theta)-\Omega_0(\theta)$ and (b) Reynolds stress $Q_{\theta\phi}$, for the mode $m = 1$ from DNS and WNL theory when $E = 1.4 \times 10^{-3}$. The blue and purple solid curves in (b) correspond to the Reynolds stress computed using $\langle u_\theta u_\phi\rangle_\phi$, and the red dot-dashed and green dotted curves corresponds when using the horizontal derivative of the differential rotation (see Eq.~(\ref{eqn:rey_chdr}), where $\nu_\mathrm{c}$ is critical viscosity) using DNS and WNL theory, respectively.}
\label{fig:rot_chng_roynolds_stress}
\end{figure}

WNL theory also predicts the strength of the Reynolds stress and its relation to the change in differential rotation. The initial rotation profile $\Omega_0(\theta)$ evolves to $\Omega(t, \theta) = \Omega_0(\theta) + \Omega_{20}(t, \theta)$, where $\Omega_{20}$ is produced due to the nonlinear interaction
of the unstable mode with its complex conjugate. In a steady-state limit, the second-order equation in Eq. (\ref{eqn:higher_order_amp_exp}) for $\psi_{20}(\theta)$ yields a balance between the Reynolds stress of the most linearly-unstable mode and the induced change in differential rotation (see Appendix \ref{app:angular_momentum} for derivation)
\begin{equation}
    Q_{11}^{\theta\phi}(\theta) := \langle u_{11}^\theta u_{11}^\phi\rangle_\phi =  \nu \sin\theta \frac{\mathrm{d}\Omega_{20}}{\mathrm{d}\theta}.
\label{eqn:rey_chdr}
\end{equation} 
The change in differential rotation and the connection with the Reynolds stress are shown in Fig.~\ref{fig:rot_chng_roynolds_stress} based on DNS and the multiscale expansion for $E= 1.4 \times 10^{-3}$. For WNL theory, we computed the Reynolds stress of all $m=1$ components that is the sum of $\psi_{11}$ and $\psi_{31}$. It explains why the Reynolds stress differs slightly from the change in differential rotation in WNL theory, but is closer to the DNS. Overall, DNS and WNL are in good agreement, even though the accuracy of the WNL approximation deteriorates as the Ekman number deviates further from the critical value. Note that the Reynolds stress in a two-dimensional model has the opposite sign compared to the observations due to the lack of latitudinal temperature difference that can be modeled in more realistic three-dimensional models \citep[see, for instance,][]{Bekki2024}.

\section{Conclusion}
We presented a simplified two-dimensional nonlinear model to explore the saturation mechanism of the linearly unstable $m = 1$ high-latitude inertial mode. In this model, the linearly unstable mode acts on the background differential rotation through its Reynolds stress (which is largest near and above the critical layer), and smooths the differential rotation profile until the mode becomes linearly stable and a  saturated state is reached. Its frequency in the saturation regime differs only slightly from its initial value owing to the nonlinearity ($\sim 1$~nHz at $E = 4 \times 10^{-4}$).
We found that the fundamental mode gives rise to harmonics with $m \geq 2$ whose amplitudes decay with $m$. These modes are known as the asymptotic Koopman modes in the literature \citep[see, for instance,][]{Bagheri2013}.  For $E = 4 \times 10^{-4}$, which corresponds to the supergranular viscosity, we found that the amplitude of the second harmonic is about eight times smaller than the fundamental, while the third harmonic is 25 times smaller. The second harmonic is always north-south antisymmetric (in vorticity) while the third harmonic has the same symmetry as the fundamental mode (north-south symmetric in this study). 

We applied two perturbations methods to model the nonlinear saturation mechanism of the $m = 1$ high-latitude mode, by performing  two different expansions of the stream-function. These two weakly nonlinear theories are valid when only one mode is unstable, as in our setup for $10^{-3} \le E < 1.5 \times 10^{-3}$. The resulting Stuart-Landau equation  gives a good description of the nonlinear saturation, with an equilibrium amplitude  proportional to  $(\Ec - E)^{1/2}$ in the multiscale expansion and to $\sigma_I^{1/2}$ in the amplitude expansion, where $\sigma_I$ is the linear growth rate. Similar relations can be found for the higher harmonics, with amplitudes that  scale as $(\Ec - E)^{m/2}$ and $\sigma_I^{m/2}$.

The implications for the interpretation of the solar observations are twofold. 
First, weakly nonlinear theory provides a quantitative framework for understanding the saturation  amplitude of the $m=1$ high-latitude mode, predicting that it scales like the square root of the linear growth rate. 
Second, some peaks in the observed power spectra may correspond to harmonics of the $m=1$ fundamental mode. Peaks at $m=2$ and $m=3$, with frequencies close to twice and three times that of the $m=1$ mode, have been reported by \citet{Gizon2021}; whether these are harmonics of the $m=1$ mode or genuine linear modes remains to be investigated.

Our results should be extended to the more realistic three-dimensional setup, where the fundamental mode also transports entropy \citep{Bekki2024}. Future work should therefore focus on analysing such three-dimensional simulations using the tools developed here, or more general frameworks for nonlinear interactions, such as Dynamic Mode Decomposition \citep[see, e.g.,][]{Schmid2022}. Another key question, which we have not addressed, concerns the pronounced modulation of the mode amplitude over the solar cycle \citep{Liang2025}.

\begin{acknowledgements}
    MM and LG thank the KAUST team for their generous hospitality, where part of this work was carried out. This project was initiated by LG, all authors performed research, MM carried out the analytical calculations, and DF the numerical simulations, and all authors contributed to writing the manuscript.  MM is a member of the IMPRS for Solar System Science at the University of G\"ottingen. LG acknowledges partial funding from ERC Synergy Grant WHOLESUN 810218 and  from the Center for Astrophysics and Space Science at the NYUAD Institute. 
\end{acknowledgements}

\bibliographystyle{aa}
\bibliography{biblio}

@ARTICLE{Crouch1993,
       author = {{Crouch}, J.~D. and {Herbert}, Th.},
        title = "{A note on the calculation of Landau constants}",
      journal = {Physics of Fluids A},
         year = 1993,
        month = jan,
       volume = {5},
       number = {1},
        pages = {283-285},
          doi = {10.1063/1.858785},
       adsurl = {https://ui.adsabs.harvard.edu/abs/1993PhFlA...5..283C}
}

@ARTICLE{Gizon2020,
       author = {{Gizon}, L. and {Fournier}, D. and {Albekioni}, M.},
        title = "{Effect of latitudinal differential rotation on solar Rossby waves: Critical layers, eigenfunctions, and momentum fluxes in the equatorial {\ensuremath{\beta}} plane}",
      journal = {\aap},
         year = 2020,
        month = oct,
       volume = {642},
          eid = {A178},
        pages = {A178},
          doi = {10.1051/0004-6361/202038525},
archivePrefix = {arXiv},
       eprint = {2008.02185},
 primaryClass = {astro-ph.SR},
       adsurl = {https://ui.adsabs.harvard.edu/abs/2020A&A...642A.178G}
}

@ARTICLE{Pham2018,
       author = {{Pham}, Khanh Gia and {Suslov}, Sergey A.},
        title = "{On the definition of Landau constants in amplitude equations away from a critical point}",
      journal = {Royal Society Open Science},
         year = 2018,
        month = nov,
       volume = {5},
       number = {11},
          eid = {180746},
        pages = {180746},
          doi = {10.1098/rsos.180746},
       adsurl = {https://ui.adsabs.harvard.edu/abs/2018RSOS....580746P}
}

@inproceedings{Landau1944,
  title={On the problem of turbulence},
  author={Landau, Lev D},
  booktitle={C.R. Acad. Sci. USSR},
  volume={44},
  pages={311},
  year={1944}
}

@ARTICLE{Stuart1960,
       author = {{Stuart}, J.~T.},
        title = "{On the non-linear mechanics of wave disturbances in stable and unstable parallel flows Part 1. The basic behaviour in plane Poiseuille flow}",
      journal = {J. Fluid Mechanics},
         year = 1960,
        month = nov,
       volume = {9},
       number = {3},
        pages = {353-370},
          doi = {10.1017/S002211206000116X},
       adsurl = {https://ui.adsabs.harvard.edu/abs/1960JFM.....9..353S}
}

@ARTICLE{Watson1960,
       author = {{Watson}, J.},
        title = "{On the non-linear mechanics of wave disturbances in stable and unstable parallel flows. Part 2. The development of a solution for plane Poiseuille flow and for plane Couette flow}",
      journal = {J. Fluid Mechanics},
         year = 1960,
        month = nov,
       volume = {9},
        pages = {371-389},
          doi = {10.1017/S0022112060001171},
       adsurl = {https://ui.adsabs.harvard.edu/abs/1960JFM.....9..371W}
}

@ARTICLE{Lindborg2022,
       author = {{Lindborg}, Erik and {Nordmark}, Arne},
        title = "{Two-dimensional turbulence on a sphere}",
      journal = {J. Fluid Mechanics},
         year = 2022,
        month = feb,
       volume = {933},
          eid = {A60},
        pages = {A60},
          doi = {10.1017/jfm.2021.1130},
       adsurl = {https://ui.adsabs.harvard.edu/abs/2022JFM...933A..60L}
}

@ARTICLE{Fournier2022,
       author = {{Fournier}, Damien and {Gizon}, Laurent and {Hyest}, Laura},
        title = "{Viscous inertial modes on a differentially rotating sphere: Comparison with solar observations}",
      journal = {\aap},
         year = 2022,
        month = aug,
       volume = {664},
          eid = {A6},
        pages = {A6},
          doi = {10.1051/0004-6361/202243473},
archivePrefix = {arXiv},
       eprint = {2204.13412},
 primaryClass = {astro-ph.SR},
       adsurl = {https://ui.adsabs.harvard.edu/abs/2022A&A...664A...6F}
}

@Book{Rudiger1989,
  title={Differential rotation and stellar convection: Sun and solar-type stars},
  author={R{\"u}diger, G{\"u}nther},
  volume={5},
  year={1989},
  publisher={Taylor \& Francis}
}

@ARTICLE{Watson1981,
       author = {{Watson}, M.},
        title = "{Shear instability of differential rotation in stars.}",
      journal = {Geophysical and Astrophysical Fluid Dynamics},
         year = 1981,
        month = jan,
       volume = {16},
       number = {4},
        pages = {285-298},
       adsurl = {https://ui.adsabs.harvard.edu/abs/1981GApFD..16..285W}
}

@ARTICLE{Cudby2021,
  title={Weakly nonlinear Holmboe waves},
  author={Cudby, Joshua and Lefauve, Adrien},
  journal={Physical Review Fluids},
  volume={6},
  number={2},
  pages={024803},
  year={2021},
  publisher={APS}
}

@ARTICLE{Dey2019,
  title={Nonlinear interaction of thermogravitational waves and thermomagnetic rolls in a vertical layer of ferrofluid placed in a normal magnetic field},
  author={Dey, Pinkee and Suslov, Sergey A},
  journal={Physics of Fluids},
  volume={31},
  number={1},
  year={2019},
  publisher={AIP Publishing}
}

@book{Schmid2012,
  title={Stability and Transition in Shear Flows},
  author={Schmid, Peter J and Henningson, Dan S},
  volume={142},
  year={2012},
  publisher={Springer Science \& Business Media}
}

@ARTICLE{Loeptien2018,
       author = {{L{\"o}ptien}, Bj{\"o}rn and {Gizon}, Laurent and {Birch}, Aaron C. and {Schou}, Jesper and {Proxauf}, Bastian and {Duvall}, Thomas L. and {Bogart}, Richard S. and {Christensen}, Ulrich R.},
        title = "{Global-scale equatorial Rossby waves as an essential component of solar internal dynamics}",
      journal = {Nature Astronomy},
         year = 2018,
        month = may,
       volume = {2},
        pages = {568-573},
          doi = {10.1038/s41550-018-0460-x},
archivePrefix = {arXiv},
       eprint = {1805.07244},
 primaryClass = {astro-ph.SR},
       adsurl = {https://ui.adsabs.harvard.edu/abs/2018NatAs...2..568L}
}

@ARTICLE{Gizon2021,
       author = {{Gizon}, Laurent and {Cameron}, Robert H. and {Bekki}, Yuto and {Birch}, Aaron C. and {Bogart}, Richard S. and {Brun}, Allan Sacha and {Damiani}, Cilia and {Fournier}, Damien and {Hyest}, Laura and {Jain}, Kiran and {Lekshmi}, B. and {Liang}, Zhi-Chao and {Proxauf}, Bastian},
        title = "{Solar inertial modes: Observations, identification, and diagnostic promise}",
      journal = {\aap},
         year = 2021,
        month = aug,
       volume = {652},
          eid = {L6},
        pages = {L6},
          doi = {10.1051/0004-6361/202141462},
archivePrefix = {arXiv},
       eprint = {2107.09499},
 primaryClass = {astro-ph.SR},
       adsurl = {https://ui.adsabs.harvard.edu/abs/2021A&A...652L...6G}
}

@ARTICLE{Hanson2022,
       author = {{Hanson}, Chris S. and {Hanasoge}, Shravan and {Sreenivasan}, Katepalli R.},
        title = "{Discovery of high-frequency retrograde vorticity waves in the Sun}",
      journal = {Nature Astronomy},
         year = 2022,
        month = mar,
       volume = {6},
        pages = {708-714},
          doi = {10.1038/s41550-022-01632-z},
       adsurl = {https://ui.adsabs.harvard.edu/abs/2022NatAs...6..708H}
}

@INPROCEEDINGS{Gizon2024,
       author = {{Gizon}, Laurent and {Bekki}, Yuto and {Birch}, Aaron C. and {Cameron}, Robert H. and {Fournier}, Damien and {Philidet}, Jordan and {Lekshmi}, B. and {Liang}, Zhi-Chao},
        title = "{Solar Inertial Modes}",
    booktitle = {IAU Symposium},
         year = 2024,
       editor = {{Getling}, Alexander V. and {Kitchatinov}, Leonid L.},
       series = {IAU Symposium},
       volume = {365},
        month = dec,
        pages = {207-221},
          doi = {10.1017/S1743921324000425},
       adsurl = {https://ui.adsabs.harvard.edu/abs/2024IAUS..365..207G}
}

@ARTICLE{Bekki2024,
       author = {{Bekki}, Yuto and {Cameron}, Robert H. and {Gizon}, Laurent},
        title = "{The Sun's differential rotation is controlled by high-latitude baroclinically unstable inertial modes}",
      journal = {Science Advances},
         year = 2024,
        month = mar,
       volume = {10},
       number = {13},
          eid = {eadk5643},
        pages = {eadk5643},
          doi = {10.1126/sciadv.adk5643},
archivePrefix = {arXiv},
       eprint = {2403.18986},
 primaryClass = {astro-ph.SR},
       adsurl = {https://ui.adsabs.harvard.edu/abs/2024SciA...10K5643B}
}

@ARTICLE{Philidet2023,
       author = {{Philidet}, J. and {Gizon}, L.},
        title = "{Interaction of solar inertial modes with turbulent convection. A 2D model for the excitation of linearly stable modes}",
      journal = {\aap},
         year = 2023,
        month = may,
       volume = {673},
          eid = {A124},
        pages = {A124},
          doi = {10.1051/0004-6361/202245666},
archivePrefix = {arXiv},
       eprint = {2304.05926},
 primaryClass = {astro-ph.SR},
       adsurl = {https://ui.adsabs.harvard.edu/abs/2023A&A...673A.124P}
}

@ARTICLE{Bekki2022a,
       author = {{Bekki}, Yuto and {Cameron}, Robert H. and {Gizon}, Laurent},
        title = "{Theory of solar oscillations in the inertial frequency range: Amplitudes of equatorial modes from a nonlinear rotating convection simulation}",
      journal = {\aap},
         year = 2022,
        month = oct,
       volume = {666},
          eid = {A135},
        pages = {A135},
          doi = {10.1051/0004-6361/202244150},
archivePrefix = {arXiv},
       eprint = {2208.11081},
 primaryClass = {astro-ph.SR},
       adsurl = {https://ui.adsabs.harvard.edu/abs/2022A&A...666A.135B}
}

@ARTICLE{Bekki2022b,
       author = {{Bekki}, Yuto and {Cameron}, Robert H. and {Gizon}, Laurent},
        title = "{Theory of solar oscillations in the inertial frequency range: Linear modes of the convection zone}",
      journal = {\aap},
         year = 2022,
        month = jun,
       volume = {662},
          eid = {A16},
        pages = {A16},
          doi = {10.1051/0004-6361/202243164},
archivePrefix = {arXiv},
       eprint = {2203.04442},
 primaryClass = {astro-ph.SR},
       adsurl = {https://ui.adsabs.harvard.edu/abs/2022A&A...662A..16B}
}

@ARTICLE{Liang2025,
  title={Doppler velocity of m= 1 high-latitude inertial mode over the last five sunspot cycles},
  author={Liang, Zhi-Chao and Gizon, Laurent},
  journal={\aap},
  volume={695},
  pages={A67},
  year={2025},
  publisher={EDP Sciences}
}

@ARTICLE{Lekshmi2026,
  title={Temporal variations of solar inertial mode parameters from GONG (2002--2024) and HMI (2010--2024): Rossby modes ($3$\backslash$leq m$\backslash$leq 16$) and $ m= 1$ high-latitude mode},
  author={Lekshmi, B and Liang, Zhi-Chao and Gizon, Laurent and Philidet, Jordan and Jain, Kiran},
  journal={arXiv preprint arXiv:2602.03741},
  year={2026}
}

@ARTICLE{Hanasoge2026,
  title={Discovery of Thermal Rossby Waves and Evidence for Weak Large-scale Convection in the Solar Interior},
  author={Hanasoge, Shravan M},
  journal={The Astrophysical Journal Letters},
  volume={997},
  number={1},
  pages={L22},
  year={2026},
  publisher={The American Astronomical Society}
}

@ARTICLE{Blume2024,
       author = {{Blume}, Catherine C. and {Hindman}, Bradley W. and {Matilsky}, Loren I.},
        title = "{Inertial Waves in a Nonlinear Simulation of the Sun's Convection Zone and Radiative Interior}",
      journal = {\apj},
         year = 2024,
        month = may,
       volume = {966},
       number = {1},
          eid = {29},
        pages = {29},
          doi = {10.3847/1538-4357/ad27d1},
archivePrefix = {arXiv},
       eprint = {2312.14270},
 primaryClass = {astro-ph.SR},
       adsurl = {https://ui.adsabs.harvard.edu/abs/2024ApJ...966...29B}
}

@ARTICLE{Rempel2005,
       author = {{Rempel}, M.},
        title = "{Solar Differential Rotation and Meridional Flow: The Role of a Subadiabatic Tachocline for the Taylor-Proudman Balance}",
      journal = {\apj},
         year = 2005,
        month = apr,
       volume = {622},
       number = {2},
        pages = {1320-1332},
          doi = {10.1086/428282},
archivePrefix = {arXiv},
       eprint = {astro-ph/0604451},
 primaryClass = {astro-ph},
       adsurl = {https://ui.adsabs.harvard.edu/abs/2005ApJ...622.1320R}
}

@ARTICLE{Boning2023,
       author = {{B{\"o}ning}, Vincent G.~A. and {Wulff}, Paula and {Dietrich}, Wieland and {Wicht}, Johannes and {Christensen}, Ulrich R.},
        title = "{Direct driving of simulated planetary jets by upscale energy transfer}",
      journal = {\aap},
         year = 2023,
        month = feb,
       volume = {670},
          eid = {A15},
        pages = {A15},
          doi = {10.1051/0004-6361/202244278},
archivePrefix = {arXiv},
       eprint = {2212.09401},
 primaryClass = {astro-ph.EP},
       adsurl = {https://ui.adsabs.harvard.edu/abs/2023A&A...670A..15B}
}

@ARTICLE{Burns1983,
       author = {{Burns}, A.~G. and {Maslowe}, S.~A.},
        title = "{Finite-Amplitude Stability of a Zonal Shear Flow.}",
      journal = {J. Atmospheric Sciences},
         year = 1983,
        month = jan,
       volume = {40},
       number = {1},
        pages = {3-9},
          doi = {10.1175/1520-0469(1983)040<0003:FASOAZ>2.0.CO;2},
       adsurl = {https://ui.adsabs.harvard.edu/abs/1983JAtS...40....3B}
}

@article{Churilov1985,
  title={Nonlinear stability of a zonal shear flow},
  author={Churilov, SM and Shukhman, IG},
  journal={Geophysical \& Astrophysical Fluid Dynamics},
  volume={36},
  number={1},
  pages={31--52},
  year={1986},
  publisher={Taylor \& Francis}
}

@ARTICLE{Larson2018,
       author = {{Larson}, Timothy P. and {Schou}, Jesper},
        title = "{Global-Mode Analysis of Full-Disk Data from the Michelson Doppler Imager and the Helioseismic and Magnetic Imager}",
      journal = {\solphys},
         year = 2018,
        month = feb,
       volume = {293},
       number = {2},
          eid = {29},
        pages = {29},
          doi = {10.1007/s11207-017-1201-5}
}

@ARTICLE{shtns,
       author = {{Schaeffer}, Nathana{\"e}L.},
        title = "{Efficient spherical harmonic transforms aimed at pseudospectral numerical simulations}",
      journal = {Geochemistry, Geophysics, Geosystems},
         year = 2013,
        month = mar,
       volume = {14},
       number = {3},
        pages = {751-758},
          doi = {10.1002/ggge.20071},
archivePrefix = {arXiv},
       eprint = {1202.6522},
 primaryClass = {physics.comp-ph},
       adsurl = {https://ui.adsabs.harvard.edu/abs/2013GGG....14..751S}
}

@article{Schmid2022,
  title={Dynamic mode decomposition and its variants},
  author={Schmid, Peter J},
  journal={Annual Review of Fluid Mechanics},
  volume={54},
  number={1},
  pages={225--254},
  year={2022},
  publisher={Annual Reviews}
}

@ARTICLE{Bhattacharya2023,
       author = {{Bhattacharya}, Jishnu and {Hanasoge}, Shravan M.},
        title = "{A Spectral Solver for Solar Inertial Waves}",
      journal = {\apjs},
         year = 2023,
        month = jan,
       volume = {264},
       number = {1},
          eid = {21},
        pages = {21},
          doi = {10.3847/1538-4365/aca09a},
archivePrefix = {arXiv},
       eprint = {2211.03323},
 primaryClass = {astro-ph.SR},
       adsurl = {https://ui.adsabs.harvard.edu/abs/2023ApJS..264...21B}
}

@ARTICLE{Mukhopadhyay2025,
       author = {{Mukhopadhyay}, Suprabha and {Bekki}, Yuto and {Zhu}, Xiaojue and {Gizon}, Laurent},
        title = "{Assessing the validity of the anelastic and Boussinesq approximations to model solar inertial modes}",
      journal = {\aap},
         year = 2025,
        month = apr,
       volume = {696},
          eid = {A160},
        pages = {A160},
          doi = {10.1051/0004-6361/202453634},
archivePrefix = {arXiv},
       eprint = {2501.16797},
 primaryClass = {astro-ph.SR},
       adsurl = {https://ui.adsabs.harvard.edu/abs/2025A&A...696A.160M}
}

@article{Durran1991,
  title={The third-order Adams-Bashforth method: An attractive alternative to leapfrog time differencing},
  author={Durran, Dale R},
  journal={Monthly weather review},
  volume={119},
  number={3},
  pages={702--720},
  year={1991}
}

@ARTICLE{Bagheri2013,
       author = {{Bagheri}, Shervin},
        title = "{Koopman-mode decomposition of the cylinder wake}",
      journal = {J. Fluid Mechanics},
         year = 2013,
        month = jul,
       volume = {726},
        pages = {596-623},
          doi = {10.1017/jfm.2013.249},
       adsurl = {https://ui.adsabs.harvard.edu/abs/2013JFM...726..596B}
}

@article{Fujimura1989,
  title={The equivalence between two perturbation methods in weakly nonlinear stability theory for parallel shear flows},
  author={Fujimura, K},
  journal={Proceedings of the Royal Society of London. A. Mathematical and Physical Sciences},
  volume={424},
  number={1867},
  pages={373--392},
  year={1989},
  publisher={The Royal Society London}
}

@article{Zhang2016,
  title={Weakly nonlinear stability analysis of subcritical electrohydrodynamic flow subject to strong unipolar injection},
  author={Zhang, M},
  journal={J. Fluid Mechanics},
  volume={792},
  pages={328--363},
  year={2016},
  publisher={Cambridge University Press}
}

@article{Herbert1983,
  title={On perturbation methods in nonlinear stability theory},
  author={Herbert, T},
  journal={J. Fluid Mechanics},
  volume={126},
  pages={167--186},
  year={1983},
  publisher={Cambridge University Press}
}

@ARTICLE{SouzaGomes2025,
       author = {{Souza-Gomes}, Mariane D. and {Finotti}, Conrado S. and {Guerrero}, Gustavo and {Triana}, Santiago A. and {Dikpati}, Mausumi and {Smolarkiewicz}, Piotr K. and {Botelho}, Eric S.},
        title = "{Non-linear simulations of the onset and non-linear dynamics of inertial waves in solar and stellar interiors}",
      journal = {arXiv e-prints},
         year = 2025,
        month = nov,
          eid = {arXiv:2511.05724},
        pages = {arXiv:2511.05724},
          doi = {10.48550/arXiv.2511.05724},
archivePrefix = {arXiv},
       eprint = {2511.05724},
 primaryClass = {astro-ph.SR},
       adsurl = {https://ui.adsabs.harvard.edu/abs/2025arXiv251105724S}
}

\appendix{}
\section{Viscous stress tensor and $\Lambda$ effect}\label{app:lambda_effect_der}
In this appendix, we derive the equation of order 0 that is used to specify the $\Lambda$ effect. As we neglected radial velocities, the divergence of the viscous tensor is given as 
\begin{align}
    \nabla \cdot \bm{\mathcal{D}} &= \frac{\nu}{r \sin \theta}\frac{\partial \mathcal{D}_{\phi\theta}}{\partial \phi}\hat{\bm{\theta}} + \frac{\nu}{r}\left[\frac{\partial \mathcal{D}_{\theta\phi}}{\partial \theta} + \cot \theta \left(\mathcal{D}_{\theta \phi} + \mathcal{D}_{\phi \theta}\right)\right]\hat{\bm{\phi}} . 
\label{eqn:div_ten}
\end{align}
Using the decomposition of the tensor given by Eq.~\eqref{eqn:viscous_tensor}, we obtain
\begin{align}
    \nabla \cdot \bm{\mathcal{D}}^\nu &= \nu \Delta \bm{u} = \nu \left(\Delta_{\rm h} + \frac{2}{r^2}\right)\bm{u} , \\
    \nabla \cdot \bm{\mathcal{D}}^\Lambda &= \frac{\nu}{r}\left(\frac{\dd }{\dd \theta}\left(\Lambda_{\theta\phi}\Omega\right) + 2\cot \theta \Lambda_{\theta \phi} \Omega\right)\hat{\bm{\phi}} .
\end{align}
At order zero, assuming that $\bm{u}_0 = r\sin\theta (\Omega_0-\Omega_\mathrm{ref}) \hat{\bm{\phi}}$, we obtain
\begin{align}
    \nabla \cdot \bm{\mathcal{D}}^\nu &= \frac{\nu}{r \sin^2 \theta} \frac{\dd}{\dd \theta}\left(\sin^3 \theta \frac{\dd \Omega_0}{\dd \theta}\right)\hat{\bm{\phi}}\\
    \nabla \cdot \bm{\mathcal{D}}^\Lambda  &= \frac{\nu}{r\sin^2\theta} \frac{\dd}{\dd\theta} \left( \sin^2\theta\Lambda_{\theta\phi}\Omega_0 \right) \hat{\bm{\phi}} .
\end{align}
In order to satisfy the equation at order zero, we need $\nabla \cdot \bm{\mathcal{D}}_0 = 0$, which gives after integration
\begin{equation}
    \sin\theta\frac{\dd \Omega_0}{\dd \theta} = -\Lambda_{\theta\phi}\Omega_0 .
\end{equation}

\section{Multiscale expansion method}
\label{app:multiscale}

\subsection{General framework}

Let us consider a nonlinear temporal evolution equation for the perturbation $q$ (most unstable mode in the system) of the form
\begin{equation}
    \frac{\partial q}{\partial t} = Q(q),
\label{eqn:gen_eqn}
\end{equation}
which we take, more specifically, as
\begin{equation}
    \frac{\partial q}{\partial t} = L q + L_\mu q + N(q,q),
    \label{eqn:gen_eqn1}
\end{equation}
where $L$ a linear operator, $L_\mu$ denotes the Jacobian with respect to a control parameter $\mu$, the term $\mathcal{N}(\cdot,\cdot)$ is a quadratic nonlinear operator, and the state vector is represented by $q(x,t)$. At a critical value $\mu = \mu_c$, the linear operator admits a nontrivial kernel associated with a marginal eigenmode.

We employ an expansion about this bifurcation point to account for finite-amplitude and saturation effects. To this end, we introduce a small parameter $\varepsilon \ll 1$, measuring the distance from criticality, and expand the solution as
\begin{equation}
    q = \varepsilon Q_1 
+ \varepsilon^2 Q_2 + \varepsilon^3 Q_3 + \cdots \quad \text{with} \quad Q_i = q_i+ q_i^* ,
\end{equation}
where $(\cdot)^*$ denotes the complex conjugate. The control parameter is expanded according to $\mu = \mu_c + \varepsilon^2 \mu_2.$ The choice of a slow time scale is crucial to ultimately obtain an evolution equation for the slow amplitude $A(T).$ The quadratic scaling can initially be replaced by a more general approach, such as $T = \varepsilon^k,$ where the exponent is determined at a later stage in the asymptotic expansion by a distinguished limit. 

An expansion at the critical point $\mu = \mu_c$ yields a hierarchy of linear problems, which can be solved in succession, and at third order in $\varepsilon$ solvability conditions will yield the desired evolution equation for $A(T).$ 
Beginning at order $\varepsilon$, we recover the linear equation 
\begin{equation}
    \frac{\partial q_1}{\partial t} = L q_1 + \mu_c L_\mu q_1 .
\label{eqn:leading_order_gen}
\end{equation}
We determine a normal modal solution $Q_1=q_1 +q_1^*$ with $q_1 = A(T)\phi_1 \mathrm{e}^{-\ii\omega t}$, where $\phi_1$ is the critical eigenfunction, $A(T)$ is a slowly varying amplitude, and the slow time scale is defined by $T=\varepsilon^2 t$. Substitution of this ansatz into the governing equation leads to an eigenvalue problem for the pair ($\phi_1, \omega)$ according to 
\begin{equation}\label{eq:linear_eig_}
    -i\omega \phi_1 = L\phi_1 + \mu_c L_\mu \phi_1 .
\end{equation}
Solving \eqref{eq:linear_eig_} produces the critical eigenfunction $\phi_1$ and its associated frequency $\omega.$
The associated adjoint eigenfunction $\phi_1^\dagger$ is defined by the corresponding adjoint problem
\begin{equation}
(+i\omega - \mathcal{L}^\dagger)\phi_1^\dagger = 0 \quad \text{with}\quad \mathcal{L}=L+\mu_c L_\mu .
\end{equation}

The next order in the asymptotic expansion, order ${\cal{O}}(\varepsilon^2),$ results in the equation 
\begin{align} \label{eq:WNL_eps2}
    \frac{\partial q_2}{\partial t} &= L q_2 + \mu_c L_\mu q_2 + \mathcal{N}(Q_1,Q_1) ,
\end{align}
The nonlinear term $\mathcal{N}(Q_1, Q_1)$ generate harmonics at frequencies $0$, $2\omega$, and $-2\omega$, allowing us to represent the particular solution as
\begin{equation}
\begin{aligned}
    q_2 \sim F_2^{(1)} A(T) e^{-i\omega t}
    &+ F_2^{(-2)} A^2(T) e^{-2i\omega t}
    + F_2^{(0)} |A(T)|^2 \\
    &+ F_2^{(2)} (A(T)^*)^2 e^{2i\omega t},
\end{aligned}
\end{equation}
where the coefficients $F_2^{(j)}$ for the particular solution at $0$, $2\omega$, and $-2\omega$ frequencies are determined by linear equations of the form $(-ij\omega -\mathcal{L}) F_2^{(j)}= f^{(j)} ),$  with $f^{(j)}$ denoting the forcing at harmonic $j$. More specifically, we obtain the system of equations 
\begin{equation}
\begin{aligned}
(-2i\omega - \mathcal{L}) F_2^{(-2)} &= \mathcal{N}(\phi_1,\phi_1), \\
(2i\omega - \mathcal{L}) F_2^{(2)} &= \mathcal{N}(\phi_1^*,\phi_1^*), \\
(-\mathcal{L}) F_2^{(0)} &= n(\phi_1,\phi_1^*)
\end{aligned}
\label{eqn:second_order_gen}
\end{equation}
for the solutions at this level of the hierarchical expansion in $\varepsilon$, where we define the symmetrized bilinear form
\begin{equation}
n(a,b) = \mathcal{N}(a,b) + \mathcal{N}(b,a).
\end{equation} 
From a physical perspective, the two solutions at this order represent the base-flow distortion by the linear perturbation and excitation of a higher-harmonic disturbance. 

At the next order, ${\cal{O}}(\varepsilon^3),$ we derive the governing equation 
\begin{align}
\frac{\partial q_3}{\partial t} + \frac{\partial q_1}{\partial T}
= & L q_3 + \mu_c L_\mu q_3  + \mu_2 L_\mu q_1 \nonumber \\ 
 & \hspace{1cm} + \mathcal{N}(Q_1, Q_2) + \mathcal{N}(Q_2, Q_1) .
\label{eq:General_eps3}
\end{align}
Upon substitution of the previous results, the right-hand side of the above equation has particular frequencies. These frequencies can be categorized into resonant and non-resonant, according to their coincidence with the frequency of the linear operator ${\mathcal{L}}$. Non-resonant terms will lead to bounded particular solutions, while resonant (or secular) terms will produce divergences and upend our hierarchical expansion approach. For this reason, we will isolate the resonant terms and project them out. Considering term by term, we can simplify the above system into 
\begin{align}
    \dfrac{\partial}{\partial t} q_3 -{\mathcal{L}}q_3 = & \mu_2L_{\mu}q_1 -\dfrac{\partial}{\partial T} q_1 \nonumber \\
    &+ n(q_1,F_2^{(0)}|A|^2)+ n(q_1^*, F_2^{(-2)} A^2 \mathrm{e}^{-2 \ii\omega t}) \nonumber \\
    &+ n(q_1^*,(F_2^{(2)}(A^*)^2 \mathrm{e}^{2 \ii\omega t)})^*)+ n(q_1, (F_2^{(0)} |A|^2)^*) \nonumber \\
    &+ \text{(non-resonant terms)}
\end{align}    
where only the critical resonant terms, proportional to $\mathrm{e}^{-\ii\omega t}\phi_1,$ have been explicitly stated on the right-hand side of the equation. In order to maintain our asymptotic expansion, these resonant terms have to be projected out by using the adjoint eigenfunction $\phi_1^\dagger$. More specifically, we require 

\begin{equation}
    \frac{\partial A}{\partial T} \langle \phi_1^\dagger,\phi_1\rangle
    = \mu_2 A(T) \langle \phi_1^\dagger, L_\mu \phi_1\rangle
      + |A|^2 A \langle \phi_1^\dagger, {\rm{SNL}} \rangle,
\end{equation}
where  ${\rm{SNL}} = n(\phi_1, F_2^{(0)})  + n(\phi_1^*, F_2^{(-2)}) + n(\phi_1,{F_2^{(0)}}^*) +n(\phi_1^*, {F_2^{(2)}}^*)$ stands for the resonant nonlinear contributions.

Finally, with the convenient normalization $\langle \phi_1^\dagger, \phi_1 \rangle=1,$ the evolution 
equation for the slow-time amplitude $A(T)$ assumes the Stuart--Landau form
\begin{equation}
    \frac{\partial A}{\partial T} = \alpha A + \Gamma |A|^2 A,
\end{equation}
with coefficients
\begin{equation}
\alpha = \mu_2 \frac{\langle \phi_1^\dagger, L_\mu \phi_1 \rangle}{\langle \phi_1^\dagger,\phi_1\rangle},
\qquad
\Gamma = \frac{\langle \phi_1^\dagger, {\rm{SNL}} \rangle}{\langle \phi_1^\dagger,\phi_1\rangle}.
\end{equation}

Before continuing with the specific terms for the solar inertial modes, a comment is necessary. The above formulation has been based on one ordering parameter, the frequency $\omega.$ This ordering parameter has been tracked through its mutual interactions towards higher harmonics, mean-flow modifications, and back to the original frequency. For more complex applications, such as the one in the following subsection, a higher-dimensional ordering parameter often arises, and the above formalism has to be extended to a parameter-tuple, in our case the (frequency, wavenumber)-pair $(\omega,m).$ Higher harmonics and conjugate tuples are defined accordingly, and the derivation of the amplitude equation proceeds analogously. 

\subsection{Derivation in our framework}

Similarly to the previous section, we measure the departure from criticality $E = \Ec + \varepsilon^2 E_2$ with $\varepsilon \ll 1$ 
and expand the stream function as
\begin{equation}
     \hat{\Psi}(\hat{T},\hat{t}, \theta, \phi) = \varepsilon\,A(\hat{T})\,\mathrm{e}^{-\ii \omega/\Omega_\mathrm{ref} \hat{t}}\Psi_1(\theta, \phi) 
    + \varepsilon^2\,q_2 + \varepsilon^3\,q_3+ \cdots + \mathrm{c.c}.
\end{equation}   
 At order $\mathcal{O}(\varepsilon)$, with $E=\Ec,$ and using the analogy to the general case \eqref{eq:linear_eig_}, we have
\begin{equation}
   -\ii \frac{\omega}{\Omega_\mathrm{ref}}(-\hat{\Delta}_{\rm h}\Psi_1)
   = (L + \Ec L_{E})\Psi_1,
   \label{eq:lin-eig-psi}
\end{equation}
where we have separated in the linear operator $\mathcal{L}$ the terms depending on the control parameters and the others $\mathcal{L} = L + \Ec L_E$.
Writing $\Psi_1(\theta, \phi) = \psi_{11}(\theta)\mathrm{e}^{\ii m \phi}$, Eq. (\ref{eq:lin-eig-psi}) can be written as
\begin{equation}
    \left(\frac{\omega}{\Omega_\mathrm{ref}} \Lm +\mathcal{L}_{m, \Ec}\right)\psi_{11}(\theta) = 0 ,
\label{eqn:leading_order}
\end{equation}
where $\hat{\omega} = \omega/\Omega_\mathrm{ref} \in \mathbb{R}$ is the (dimensionless) frequency of the critical eigen-mode $\psi_{11}(\theta)$ and the linear operators $\Lm$ and $\mathcal{L}_{m, \Ec}$ are defined in Eqs.~\eqref{eqn:L_m} and \eqref{eqn:L_A} respectively.
Equation~\eqref{eqn:leading_order} is the linear eigenvalue problem at $\Ec$ considered in \citep[][]{Fournier2022} which determines the critical eigen-pair $(\psi_{11}, \omega/\Omega_\mathrm{ref})$. 

The adjoint eigenfunction $\psi_{11}^\dagger$, which will be used at order ${\cal{O}}(\varepsilon^3),$ is defined by the associated adjoint problem:
\begin{equation}
    \left(\frac{\omega}{\Omega_\mathrm{ref}}\mathbb{L}_{-m} + \mathcal{L}^\dagger_{m, \Ec} \right) \psi^\dagger_{11}(\theta) = 0 ,
\label{eqn:Adjoint}
\end{equation}

At order ${\cal{O}}(\varepsilon^2)$ the base flow is modified by  nonlinear interactions of the perturbations  and higher harmonics are generated. With our notations, Eq.~\eqref{eq:WNL_eps2} is written as
\begin{equation}
   \frac{\partial}{\partial \hat{t}}\big(-\hat{\Delta}_{\rm h}q_2\big) = (L + \Ec\,L_E)q_2 + \mathcal{N}[Q_1, Q_1] ,
\label{eqn:order2}
\end{equation}
where $Q_1 = q_1 + q_1^*$ with $q_1 = A(\hat{T})\psi_{11}(\theta)\mathrm{e}^{-\ii(\omega/\Omega_\mathrm{ref}\hat{t} - m \phi)}$. The quadratic nonlinear term $\mathcal{N}[Q_1, Q_1]$ creates harmonics at $(\omega=0,m=0)$, $(2\omega,2m)$ and $(-2\omega,-2m)$ and we seek for a solution of the form
\begin{equation}
    q_2 = |A(\hat{T})|^2 \psi_{20}(\theta) + \left(A^2(\hat{T}) \psi_{22}(\theta) \mathrm{e}^{-2 \ii(\omega/\Omega_\mathrm{ref}\hat{t} -m \phi)} + \mathrm{c.c.}\right) .
\label{eqn:q211}
\end{equation}
Employing this, Eq. (\ref{eqn:order2}) gives rise to the system of equations
\begin{align}
    \mathcal{L}_{0, \Ec} \psi_{20}(\theta) &= \ii f_{20}(\theta) \, , \label{eqn:2nd1} \\
    \left(\frac{2\omega}{\Omega_\mathrm{ref}} \mathbb{L}_{2m} +  \mathcal{L}_{2m, \Ec} \right) \psi_{22}(\theta) &= \ii f_{22}(\theta) \, , \label{eqn:2nd2}
\end{align}
where $f_{20}$ and $f_{22}$ collect the terms from $\mathcal{N}[Q_1, Q_1]$ that corresponds to $(\omega=0,m=0)$ and $(2\omega,2m)$ respectively and are given as
\begin{align}
    f_{20} &=  \frac{\ii m}{\sin\theta} \left( \psi^\prime_{11} \overline{\zeta_{11}} + \psi_{11} \overline{\zeta^\prime_{11}} - \overline{\psi^\prime_{11}}  \zeta_{11} - \overline{\psi_{11}} \zeta^\prime_{11}  \right) , \label{eqn:f20} \\
    f_{22} &= \frac{\ii m}{ \sin\theta} \left(\psi_{11} \zeta^\prime_{11}  - \psi^\prime_{11} \zeta_{11}\right) , \label{eqn:f22}
\end{align}
where prime denote the derivative with respect to colatitude and overline represents the complex conjugate and $\zeta_{11} := -\Lm \psi_{11}$. 

Finally, at the next order ${\cal{O}}(\varepsilon^3),$ the slow time-scale enters the asymptotic expansion 
via $\partial_{\hat{t}}\to \partial_{\hat{t}} + \varepsilon^2 \partial_{\hat{T}},$ and yields an evolution equation for the slowly varying amplitude. The full equations at this stage reads 
\begin{align}
    \frac{\partial}{\partial \hat{t}} (-\hat{\Delta}_\mathrm{h}q_3) + \frac{\partial}{\partial \hat{T}}(-\hat{\Delta}_\mathrm{h} q_1) &= \left(L + \Ec L_{E}\right)q_3 + E_2 L_{E} q_1  \nonumber \\
    &\hspace{0.5cm}+ \mathcal{N}[Q_1, Q_2] + \mathcal{N}[Q_2, Q_1] \, ,
\label{eqn:order_eps3}
\end{align}
where $Q_j = q_j+q_j^*$. Using the single-mode ansatz $q_1$ and the second-order particular solution (\ref{eqn:q211}), the third-order equation (\ref{eqn:order_eps3}) can be written as
\begin{align}
    &\frac{\partial}{\partial \hat{t}} (-\Delta_\mathrm{h}q_3) + \left(\frac{\partial A}{\partial \hat{T}} \zeta_{11}\mathrm{e}^{-\ii(\omega/\Omega_\mathrm{ref} \hat{t} - m \phi)} + \mathrm{c.c}\right) = \left(L + \Ec L_{E}\right)q_3 \nonumber \\ 
    &\hspace{1.5cm} + \left(E_2(\Lm + 2)\zeta_{11} A \mathrm{e}^{-\ii(\omega/\Omega_\mathrm{ref} \hat{t} - m \phi)} + \mathrm{c.c}\right) \nonumber \\
    & \hspace{1.5cm} -\left(|A(\hat{T})|^2A(\hat{T})\,\mathrm{e}^{-\ii(\omega/\Omega_\mathrm{ref} \hat{t} - m\phi)} f_{31}(\theta) \nonumber \right. \\ 
    &\hspace{1.5cm} \left. + A^3(\hat{T}) \mathrm{e}^{-3 \ii(\omega/\Omega_\mathrm{ref} \hat{t} -m \phi)}f_{33}(\theta) + \mathrm{c.c}\right) ,
\label{eqn:order_eps3a}
\end{align}
where $f_{31}(\theta)$ and $f_{33}(\theta)$ are given as
\begin{align}
    f_{31} &= \frac{\ii m }{\sin \theta}\left(\psi_{22}^\prime \overline{\zeta_{11}} + \zeta_{20}^\prime \psi_{11} + 2\psi_{22} \overline{\zeta_{11}^\prime} - \zeta_{11} \psi_{20}^\prime \right. \nonumber \\
    &\left.\hspace{4cm} - 2\zeta_{22} \overline{\psi_{11}^\prime} - \zeta_{22}^\prime \overline{\psi_{11}}\right) , \label{eqn:f31} \\
    f_{33} &= \frac{\ii m }{\sin \theta}\left(2 \zeta_{11}^\prime \psi_{22} + \zeta_{22}^\prime \psi_{11} - \zeta_{11} \psi_{22}^\prime - 2 \zeta_{22} \psi_{11}^\prime\right) , \label{eqn:f33} 
\end{align}
and we have defined $\zeta_{m'm''}(\theta) := -\mathbb{L}_{m''}\psi_{m'm''}(\theta)$.
We look for a solution $q_3$ of Eq.~\eqref{eqn:order_eps3a} as
\begin{align}
    q_3(\hat{T}, \hat{t}, \theta, \phi) &= |A(\hat{T})|^2A(\hat{T}) \mathrm{e}^{-\ii(\omega/\Omega_\mathrm{ref}\hat{t} - m \phi)} \psi_{31}(\theta) \nonumber \\ 
    &\hspace{1cm}+ A^3(\hat{T}) \psi_{33}(\theta) \mathrm{e}^{-3\ii(\omega/\Omega_\mathrm{ref}\hat{t} -m \phi)} + \mathrm{c.c} .
\label{eqn:q311}
\end{align}
Using the third-order perturbation (\ref{eqn:q311}) into Eq. (\ref{eqn:order_eps3a}) and collecting the coefficients of $\mathrm{e}^{-3 \ii(\hat{\omega}\hat{t} - m\phi)}$ and $\mathrm{e}^{-\ii(\omega/\Omega_\mathrm{ref}\hat{t}-m\phi)}$ with third order in amplitude $A(\hat{T})$, yields the two third-order equations, for the third harmonic and the fundamental correction, which are given as
\begin{align}
    \left(\frac{3\omega}{\Omega_\mathrm{ref}}\mathbb{L}_{3m} + \mathcal{L}_{3m,\Ec} \right)\psi_{33}(\theta) &= \ii f_{33}(\theta) , \label{eqn:3rd1} \\
    |A|^2A \left(\frac{\omega}{\Omega_\mathrm{ref}}\Lm + \mathcal{L}_{m,\Ec} \right)\psi_{31}(\theta) &= \ii |A|^2A f_{31}(\theta) + \ii \frac{\partial A}{\partial \hat{T}}\zeta_{11}(\theta) \nonumber \\ 
    &\hspace{0.25cm} - \ii A E_2\left(\Lm + 2\right)\zeta_{11}(\theta) . \label{eqn:3rd2}
\end{align}
The linear operator on left-hand side in Eq. (\ref{eqn:3rd2}) is non-singular and hence always gives a unique solution $\psi_{33}(\theta)$. The linear operator acting on $\psi_{31}$ on the left-hand side of Eq. (\ref{eqn:3rd2}) is identical to the operator governing the critical eigenmode and is therefore singular by construction. As a consequence, Eq. (\ref{eqn:3rd2}) does not admit a solution for $\psi_{31}$ unless its right-hand side satisfies a compatibility condition. According to the Fredholm alternative theorem, this condition requires that the forcing term on the right-hand side be orthogonal to the null space of the adjoint operator.

We therefore take the inner product of Eq. (\ref{eqn:3rd1}) with the adjoint eigenfunction $\psi_{11}^\dagger$, which spans the kernel of the adjoint operator, in order to enforce this solvability condition. This procedure eliminates the resonant contribution proportional to the critical mode and yields a closed evolution equation for the slowly varying amplitude $A(\hat{T})$, famously called the cubic \textit{Stuart-Landau equation} \citep{Landau1944, Stuart1960}  
\begin{equation}
    \frac{\partial A}{\partial \hat{T}} = -\ii \alpha A - \ii\Gamma |A|^2A ,
\label{eqn:Stuart_Landau_eq11}
\end{equation}
where $\sigma$ and $\beta$ are called the Landau coefficients which are defined as
\begin{align}
    \alpha := E_2 \xi := E_2\frac{\langle \ii(\Lm +2)\zeta_{11}, \psi_{11}^\dagger\rangle}{\langle\zeta_{11}, \psi_{11}^\dagger\rangle} , \quad
    \Gamma := -\frac{\langle \ii f_{31}, \psi_{11}^\dagger\rangle}{\langle\zeta_{11}, \psi_{11}^\dagger\rangle} .
\label{eqn:sigma_beta}
\end{align}
Substituting Eq. (\ref{eqn:Stuart_Landau_eq11}) into Eq. (\ref{eqn:3rd2}) and employing the gauge condition, $\langle \psi_{31}, \psi_{11}^\dagger\rangle = 0$, all terms proportional to $\zeta_{11}$ vanish and we obtain (see proof in Appendix~\ref{sect:app_psi31})
\begin{equation}
    \left(\frac{\omega}{\Omega_\mathrm{ref}}\Lm + \mathcal{L}_{m,\Ec} \right)\psi_{31}(\theta) = \ii f_{31}(\theta) + \Gamma \zeta_{11}(\theta) .
\label{eqn:3rd2a}
\end{equation}
The solvability condition embodied in the amplitude equation removes the resonant forcing from Eq. (\ref{eqn:3rd2}). The resulting Eq. (\ref{eqn:3rd2a}) is therefore no longer singular, since its right-hand side is orthogonal to the adjoint null space of the linear operator. The operator acting on $\psi_{31}$ can thus be inverted, yielding a unique solution for $\psi_{31}$ that describes the nonlinear correction to the fundamental eigenfunction.

\subsection{Derivation of Eq.~\eqref{eqn:3rd2a}}
\label{sect:app_psi31}

 Substituting Eq. (\ref{eqn:Stuart_Landau_eq11}) into Eq. (\ref{eqn:3rd2}), we obtain
\begin{align}
    \left(\hat{\omega}\Lm + \mathcal{L}_{m,\Ec} \right)\psi_{31}(\theta) &= \ii f_{31}(\theta) + \Gamma \zeta_{11}(\theta)  \nonumber \\
    & +\frac{1}{|A|^2}\left[\alpha - \ii E_2\left(\Lm + 2\right)\right]\zeta_{11}(\theta) \, 
    .
\end{align}
Since $(\hat{\omega}\Lm + \mathcal{L}_{m, \Ec})\psi_{11} = 0$ which implies $(\hat{\omega}\Lm + \mathcal{L}_{m, \Ec})\zeta_{11} = 0$, that is, $\zeta_{11}$ is also the right null vector of the linear operator $(\hat{\omega}\Lm + \mathcal{L}_{m, \Ec})$ up to some constant factor as $\zeta_{11} = -\Lm\psi_{11}$. That means that any term proportional to $\zeta_{11}$ is resonant, falls into the null-space of this operator and, thus, the equation cannot be solved directly. Hence, Eq. (\ref{eqn:3rd2a}) has no unique solution $\psi_{31}$ because, if  $\psi_{31}$ is a solution, so is the $\psi_{31} + C \psi_{11}$, where $C$ as non-zero complex constant. To remedy this non-uniqueness, we must remove the components parallel to the null vector $\zeta_{11}$ which can be achieved by invoking the gauge condition (normalization). We define the projection operator
\begin{equation}
    P = I - \zeta_{11}\frac{\langle \cdot, \psi_{11}^\dagger\rangle}{\langle \zeta_{11}, \psi_{11}^\dagger\rangle} \, ,
\end{equation}
where $I$ is an identity operator. This operator enforces the orthogonality of $\psi_{31}(\theta)$ to the adjoint of the neutral mode $\psi_{11}^\dagger(\theta)$. Hence the projection operator contains only the non-resonant spatial structures that can actually be inverted by the linear operator $(-\ii\hat{\omega}-\mathcal{L}_{m, \Ec})$. Applying the projection operator to Eq. (\ref{eqn:3rd2a}) on both sides and using $P \zeta_{11} = 0$, we get
\begin{align}
    P\left((\hat{\omega}\Lm + \mathcal{L}_{m,\Ec})\psi_{31}\right) &= P\left(\ii f_{31}\right) . \label{eqn:3rd2b}
\end{align}
Employing the gauge condition, $\langle \psi_{31}, \psi_{11}^\dagger\rangle = 0$, all terms proportional to $\zeta_{11}$ vanish, yielding the solvable boundary-value problem for the fundamental correction $\psi_{31}$, with same boundary conditions as in leading-order equation
\begin{equation}
    (\hat{\omega}\Lm + \mathcal{L}_{m,\Ec})\psi_{31} = \ii f_{31} - \zeta_{11}\frac{\langle \ii f_{31}, \psi_{11}^\dagger\rangle}{\langle \zeta_{11}, \psi_{11}^\dagger\rangle} , \label{eqn:3rd2c}
\end{equation}
which is same as Eq. (\ref{eqn:3rd2a}) after using the definition of the second Landau coefficient from Eq. (\ref{eqn:sigma_beta}).
Since $\Gamma$ is determined uniquely in Eq. (\ref{eqn:sigma_beta}), the spatial distribution of the fundamental correction $\psi_{31}$ can be uniquely determined from Eq. (\ref{eqn:3rd2a}). 

\section{Amplitude expansion method}\label{app:amplitude_expansion}

\subsection{General framework}

In the amplitude expansion method, the time-dependent disturbance amplitude $A(t)$ itself serves as a small parameter. Following this, we do not need to separate the different time scales and hence no need to define the detuning coefficient $\mu_2$ in the control parameter. 

Assuming only one unstable mode, we write the expansion of this mode as
\begin{equation}
    q_1(x, t) = A(t)\phi^{(1)}_1(x) + \mathrm{c.c}.
\label{eqn:q1_exp}
\end{equation}
Substituting Eq. (\ref{eqn:q1_exp}) expansion into Eq. (\ref{eqn:gen_eqn1}), we obtain the leading-order equation at $\mathcal{O}(A)$ similar to Eq. (\ref{eqn:leading_order_gen}) which has to be solved at the considered parameter $\mu$, and not at the critical parameter $\mu_c$. From this leading order equation we obtain the complex eigenvalue $\sigma = \sigma_R+i\sigma_I$, with $\sigma_R$ as the frequency and $\sigma_I$ the linear growth rate of the mode. This linear eigenvalue problem is satisfied for any time if we assume
\begin{equation}
    \frac{\mathrm{d} A(t)}{\mathrm{d}t} = -i \sigma A(t) .
\label{eqn:lindiff}
\end{equation}
The quadratic nonlinear term $\mathcal{N}(q, q)$ give terms like $|A(t)|^2F^{(0)}_2$ and $A^2(t)F_2^{(2)}$, where $F_2^{(0)}$ and $F_2^{(2)}$ depends only on the spatial structure of the linear eigenfunction $\phi^{(1)}_1$ at the considered control parameter. To compensate for these terms we add the similar terms in the expansion (\ref{eqn:q1_exp}) such that
\begin{equation}
    q_1(x, t) = |A(t)|^2\phi^{(0)}_{2}(x) + \left(A(t)\phi^{(1)}_1(x) + A^2(t) \phi_2^{(2)}(x) + \mathrm{c.c}\right) .
\label{eqn:q1_exp11}
\end{equation}
On substituting this expansion into Eq. (\ref{eqn:gen_eqn1}) and equating the terms of $\mathcal{O}(|A(t)|^2)$ and $\mathcal{O}(A^2(t))$ yields the two (inhomogeneous linear) equations for the base-flow distortion and the second harmonic, similar to Eq. (\ref{eqn:second_order_gen}), but unlike in multiscale expansion, the equations has to be solved at considered control parameter $\mu$. The linear operators on the left-hand side of these second-order equations are non-singular and hence always have unique solutions. 

Substituting Eq. (\ref{eqn:q1_exp11}) into Eq. (\ref{eqn:gen_eqn1}), the nonlinear term $\mathcal{N}(q_1, q_1)$ gives terms like $|A(t)|^2 A(t)F_{3}^{(1)}$ and $A^3(t)F_3^{(3)}$, where $F_{3}^{(1)}$ and $F_{3}^{(3)}$ depends on $\phi^{(1)}_1$, $\phi_2^{(0)}$ and $\phi_2^{(2)}$. To compensate for these terms, we add similar terms in the expansion (\ref{eqn:q1_exp11})
\begin{align}
    q_1(x, t) &= |A(t)|^2\phi^{(0)}_{2}(x) + \left(A(t)\phi^{(1)}_1(x) + |A(t)|^2A(t)\phi^{(1)}_3(x) \nonumber \right.\\ 
    &\hspace{1.5cm}\left.+ A^2(t) \phi_2^{(2)}(x) + A^3(t) \phi_3^{(3)}(x) + \mathrm{c.c}\right) .
\label{eqn:q1_exp22}
\end{align}
Substituting Eq. (\ref{eqn:q1_exp22}) into Eq. (\ref{eqn:gen_eqn1}) and equating the coefficients of $\mathcal{O}(|A(t)|^2A(t))$ and $\mathcal{O}(A^3(t))$, we get the two third-order equations similar as in multiscale expansion, for the correction to the fundamental and the third harmonic which has to be solved at considered control parameter $\mu$, not just at $\mu_c$ as in multiscale expansion method.

At third order, the equation governing the correction $\phi_3^{(1)}$ to the fundamental mode contains resonant terms proportional to $\phi_1^{(1)}$. In the limit where $\sigma_I \to 0$, the associated linear operator becomes singular and a solvability condition must be imposed to avoid secular growth. We therefore assume that the time-dependent amplitude satisfies a cubic Stuart–Landau equation
\begin{equation}
    \frac{\mathrm{d} A(t)}{\mathrm{d}t} = -i \sigma A(t) - i \beta |A(t)|^2A(t) ,
\label{eqn:nonlindiff}
\end{equation}
where $\sigma$ is the linear eigenvalue of the most unstable mode at the considered control parameter, and $\beta$ is a nonlinear Landau coefficient to be determined. 

When employing Eq. (\ref{eqn:nonlindiff}) instead of Eq. (\ref{eqn:lindiff}), we obtain the modified inhomogeneous equation for the fundamental correction $\phi_3^{(1)}$. For $\sigma_I = 0$ (that is at $\mu_\mathrm{c}$), the second Landau coefficient can be uniquely determined by using the solvability condition, however its value is not uniquely determined for $\sigma_I \neq 0$ since the modified equation is solvable for any value of $\beta$ in that case. \citet{Crouch1993} proposed to impose the weighted orthogonality condition between the fundamental mode $\phi_1^{(1)}$ and the nonlinear correction $\phi_3^{(1)}$ which gets rid of the resonant terms. On combining this weighted orthogonality condition with the modified inhomogeneous equation for the fundamental correction, we obtain the augmented system which at a given parameter value $\mu$ can be solved for both the fundamental correction $\phi_3^{(1)}$ and the second Landau coefficient $\beta$ uniquely.

\subsection{Derivation in our framework}

We write the normalized perturbation of the stream function $\hat{\Psi}$ as
\begin{equation}
    \hat{\Psi}(\hat{t},\theta,\phi) =  A(\hat{t}) \psi_{11}(\theta) \mathrm{e}^{\ii m\phi} + \textrm{c.c.} . 
\label{eqn:psi11a}
\end{equation}
The radial vorticity is then given as $\hat{Z} = -A(\hat{t})\mathbb{L}_m\psi_{11}(\theta)e^{im\phi} + \textrm{c.c}$,
with $\mathbb{L}_m$ defined in Eq. (\ref{eqn:L_m}).

Upon substituting stream function expansion (\ref{eqn:psi11a}) into Eq. (\ref{eqn:psi_eqn}), we obtain
\begin{align}
    &\left(\mathbb{L}_m \psi_{11}(\theta) \frac{\dd A(\hat{t})}{\dd \hat{t}} - \ii \mathcal{L}_{m, E}\psi_{11}(\theta) A (\hat{t}) \right) \mathrm{e}^{\ii m \phi} + \textrm{c.c.} \nonumber\\
   &\hspace{2cm} = |A(\hat{t})|^2 f_{20}(\theta) + \left(A(\hat{t})^2 f_{22}(\theta) \mathrm{e}^{2 \ii m \phi} + \textrm{c.c.}\right) , 
\label{eqn:amp1b}
\end{align}
where the linear operators $\mathbb{L}_m$ and $\mathcal{L}_{m, E}$ are defined in Eqs. (\ref{eqn:L_m}, \ref{eqn:L_A}) and the expressions for nonlinear functions $f_{20}(\theta)$ and $f_{22}(\theta)$ are the same as given in Eqs. (\ref{eqn:f20}) and (\ref{eqn:f22}). Please note that in this frame work, $f_{20}$ and $f_{22}$ are not computed at the critical point, but at the considered Ekman number.

Using the wave ansatz $A(\hat{t}) = A_0 \mathrm{e}^{-\ii \sigma \hat{t}/\Omega_\mathrm{ref}}$, with $\sigma$ as complex and $A_0 = A(\hat{t} = 0)$, the linear terms in amplitude $A(\hat{t})$ give
\begin{equation}
    \left(\frac{\sigma}{\Omega_\mathrm{ref}}\mathbb{L}_m + \mathcal{L}_{m, E}\right) \psi_{11}(\theta) = 0 .
\label{eqn:lin_prob}
\end{equation}
This corresponds to the linear problem considered by \citet{Fournier2022} and reduces to \citeauthor{Watson1981}'s  (\citeyear{Watson1981}, their equation 2.10) in the inviscid case.

Due to the nonlinear term, quantities in $A^2(\hat{t})$ and $|A(\hat{t})|^2$ appear at second-order in the amplitude disturbance, see right-hand side of Eq.~\eqref{eqn:amp1b}. To compensate for these terms, we need to add similar terms in the stream function expansion (\ref{eqn:psi11a})
\begin{align}
    \hat{\Psi}(\hat{t},\theta,\phi) = & |A(\hat{t})|^2 \psi_{20}(\theta) \nonumber \\
    & + \left( A(\hat{t}) \psi_{11}(\theta) \mathrm{e}^{\ii m\phi}  
  + A^2(\hat{t}) \psi_{22}(\theta) \mathrm{e}^{2 \ii m \phi} +  \textrm{c.c.} \right) . 
\label{eqn:psi22}
\end{align}
The background flow is modified by the action of the mode and becomes $\psi_{00}(\theta) + |A(\hat{t})|^2 \psi_{20}(\theta)$. Moreover, the second harmonic of the unstable mode $\psi_{22}(\theta)$ appears at an azimuthal order $2m$. Substituting Eq.~\eqref{eqn:psi22} into Eq.~\eqref{eqn:psi_eqn} and equating the terms independent of $\phi$  and those in $\mathrm{e}^{2 i m \phi}$ respectively gives 
\begin{align}
& \left(\frac{2i\sigma_I}{\Omega_\mathrm{ref}}\mathbb{L}_0 + \mathcal{L}_{0,E}\right)\psi_{20} = i f_{20} , \label{eqn:2nd_order1a} \\
   &  \left(\frac{2\sigma}{\Omega_\mathrm{ref}}\mathbb{L}_{2m} + \mathcal{L}_{2m,E}\right)\psi_{22} = i f_{22} , \label{eqn:2nd_order2a}
\end{align}
where $\sigma_I = \mathrm{Im}( \sigma)$ and the linear operators $\mathbb{L}_m$ and $\mathcal{L}_{m,\sigma}$ was defined earlier (see Eq.~\eqref{eqn:L_A}).
The linear operators on the left-hand sides are non-singular, and thus invertible even in the limit $\sigma_I \to 0$ \citep[see, for instance,][]{Crouch1993}. 

When substituting Eq.~(\ref{eqn:psi22}) into the nonlinear term $\mathcal{N}$ of Eq.~(\ref{eqn:psi_eqn}), two extra terms, $f_{31}(\theta) |A(\hat{t})|^2 A(\hat{t}) \mathrm{e}^{\ii m \phi}$ and $f_{33} A(\hat{t})^3 \mathrm{e}^{3 \ii m \phi}$ appear. To treat these terms, we need to add similar terms to the normalized stream function $\hat{\Psi}$ in Eq.~(\ref{eqn:psi22}) and look for a solution of the form
\begin{align}
    \hat{\Psi}(\hat{t},\theta,\phi) = & |A(\hat{t})|^2 \psi_{20}(\theta)
    + \left(A(\hat{t}) \psi_{11}(\theta) \mathrm{e}^{\ii m\phi} \right. \nonumber \\
    &\left. + A^2(\hat{t}) \psi_{22}(\theta) \mathrm{e}^{2 \ii m \phi} + |A(\hat{t})|^2 A(\hat{t}) \psi_{31}(\theta) \mathrm{e}^{\ii m\phi} \right. \nonumber \\
    &\left. + \psi_{33}(\theta) A^3(\hat{t}) \mathrm{e}^{3 \ii m \phi} + \textrm{c.c.} \right) .
\end{align}
Substituting into Eq. (\ref{eqn:forward_norm}) and collecting the terms in $|A(\hat{t})|^2A(\hat{t}) e^{i m \phi}$ and $A(\hat{t})^3 \mathrm{e}^{3 \ii m\phi}$ respectively gives
\begin{align}
   & \left(\frac{\sigma+2i\sigma_I}{\Omega_\mathrm{ref}}\mathbb{L}_m + \mathcal{L}_{m,E}\right)\psi_{31} = i f_{31} , \label{eqn:3rd_order1a} \\
    & \left(\frac{3\sigma}{\Omega_\mathrm{ref}}\mathbb{L}_{3m} + \mathcal{L}_{3m,E}\right)\psi_{33} = i f_{33} , \label{eqn:3rd_order2a}
\end{align}
where the expressions of right-hand sides $f_{31}(\theta)$ and $f_{33}(\theta)$ are similar to the ones given in Eqs. (\ref{eqn:f31}) and (\ref{eqn:f33}) but computed at the considered Ekman number instead of the critical one.

If $\sigma_I > 0$, then left-hand sides of Eqs. (\ref{eqn:3rd_order1a}) and (\ref{eqn:3rd_order2a}) are non-singular and hence the unique solutions $\psi_{31}(\theta)$ and $\psi_{33}(\theta)$ can be found. However, when $\sigma_I \to 0$, then the linear operator on left-hand side of Eq. (\ref{eqn:3rd_order1a}) is singular by construction \citep[see, for instance,][]{Crouch1993}. To resolve this potential unsolvability problem for function $\psi_{31}(\theta)$, we need to assume the time evolution of the disturbance amplitude can not be determined from linear theory alone, instead we need to add one more term in the evolution equation as follows
\begin{equation}
    \frac{\mathrm{d} A(\hat{t})}{\mathrm{d} \hat{t}} = -\ii \left( \frac{\sigma}{\Omega_\mathrm{ref}} A(\hat{t}) + \frac{\beta}{\Omega_\mathrm{ref}} |A(\hat{t})|^2 A(\hat{t}) \right) .
\label{eqn:cubic_SL_eqn1}
\end{equation}
This is the Stuart-Landau equation \citep[see, e.g.,][]{Landau1944, Stuart1960}, where $\sigma$, the first Landau coefficient, determines the linear growth rate and $\beta$ is the second Landau constant which are yet to be determined. Both coefficients have units of frequencies as we are working with dimensionless quantities.
Using Eq.~\eqref{eqn:cubic_SL_eqn1} instead of wave ansatz, affects only the equation for $\psi_{31}(\theta)$ which becomes
\begin{equation}
    \left(\frac{\sigma+2i\sigma_I}{\Omega_\mathrm{ref}}\mathbb{L}_m + \mathcal{L}_{m, E}\right) \psi_{31}(\theta) = \ii f_{31}(\theta) - \frac{\beta}{\Omega_\mathrm{ref}} \mathbb{L}_m \psi_{11}(\theta) .
\label{eqn:f31_landau}
\end{equation}
In the resonant case ($\sigma_I = 0$), the second Landau constant is uniquely defined and can be obtained by using the Fredholhm-alternative theorem. However, when $\sigma_I \neq 0$,  the value of the Landau constant is not uniquely determined since the inhomogeneous system (\ref{eqn:f31_landau}) is solvable for any value of $\beta$ and an additional constrain has to be added. \citet{Crouch1993} propose to assume weighted orthogonality between $\psi_{11}$ and $\psi_{31}$, that is,
\begin{equation}
    \langle \psi_{11} , \, \psi_{31}\rangle_\mathcal{M} := \psi_{11}^H\ \mathcal{M} \psi_{31} = 0 ,
\label{eqn:extra_cond11}
\end{equation}
where $\mathcal{M}$ is a positive definite matrix and $(\cdot)^H$ denotes the Hermitian transpose.  \citet{Pham2018} showed that this condition is equivalent to the solvability condition when $\sigma_I \to 0$.  Several choices are possible for $\mathcal{M}$ \citep[][their section~4]{Pham2018}. Here, we choose the simplest option: $\mathcal{M} = \mathbb{I}$, the identity matrix. Then the Landau constant $\beta$ and the function $\psi_{31}$ are thus determined by solving the extended system
\begin{equation}
    \tilde{\mathcal{L}} \tilde{\psi}_{31} = \ii \tilde{f}_{31} , \quad     \tilde{\psi}_{31} = \left[ \begin{array}{c} \psi_{31} \\ \beta/\Omega_\mathrm{ref} \end{array} \right] , \quad 
    \tilde{f}_{31} = \left[ \begin{array}{c}  f_{31}  \\ 0 \end{array} \right]  , 
\label{eqn:augmented_sys}
\end{equation}
where
\begin{equation}
    \tilde{\mathcal{L}} = \left[
    \begin{array}{cc}
    (\sigma+2i\sigma_I)/\Omega_\mathrm{ref} \mathbb{L}_m +\mathcal{L}_{m, E} &  \mathbb{L}_m \psi_{11} \\
    \psi_{11}^H \mathcal{M} & 0
    \end{array}
    \right] .
\end{equation}

\section{Dependence of Landau coefficients on Ekman number}
\begin{figure}[hbt!]
    \centering
    \includegraphics[width=0.9\linewidth]{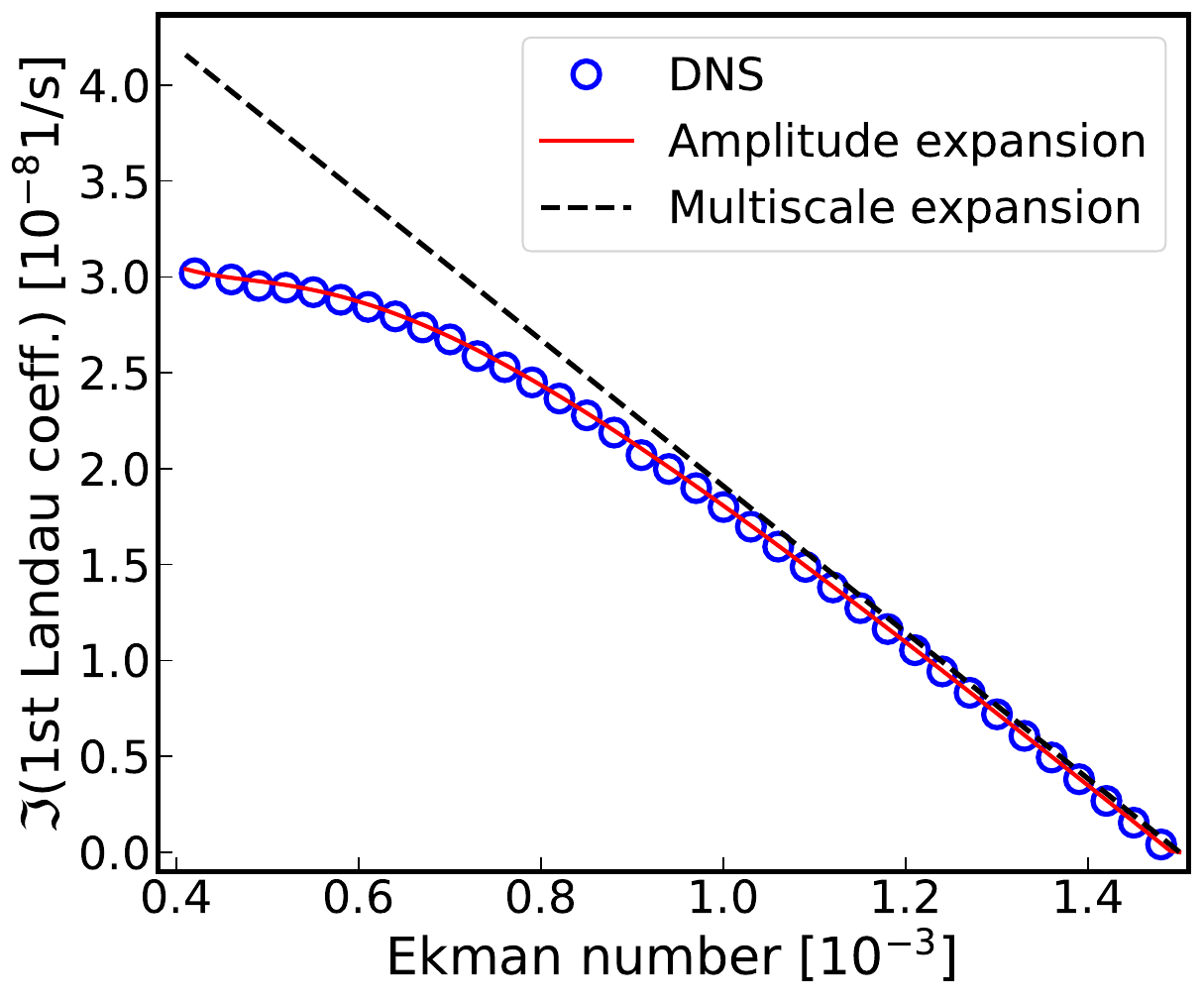}
\caption{The imaginary part of the first Landau coefficient obtained by fitting the DNS data (see Eq.~\eqref{eqn:lin_func_uavg}) and from WNL theory for the most unstable $m = 1$ mode.}
\label{fig:growth_rate}
\end{figure}

Figure~\ref{fig:growth_rate}  compares the linear Landau coefficient computed using the two weakly nonlinear expansion methods to the one obtained by fitting the Stuart-Landau model to the DNS data.
As expected, both agree very well close to the critical Ekman number and deviate progressively more as $\Ec-E$ increases. Note that this approximation is not valid for $E < 10^{-3}$ as multiple modes become unstable.  
\begin{table*}[t]
\caption{Ekman number dependence of imaginary parts of the Landau coefficients for the $m = 1$ mode.}
\centering
\begin{tabular}{ccccccc}
\hline
\hline
$E \, [10^{-3}]$ & \multicolumn{3}{c}{$\Im$(1st Landau coeff.) $ [10^{-9}\mathrm{s}^{-1}]$}  & \multicolumn{3}{c}{$\Im$(2nd Landau coeff.) $[10^{-11} \mathrm{s/m}^2]$}           \\
& DNS & Amp.exp. & Mul. exp. & DNS & Amp. exp. & Mul. exp. \\
\hline
$1.48$       &  $0.40$    & 0.45   & $0.77$   & $-3.54$   &     $-3.54$   & $-3.17$   \\
$1.40$       &  $3.43$    & 3.47   & $3.82$   & $-3.66$   &     $-3.66$   & --        \\
$1.30$       &  $7.19$    & 7.23   & $7.64$   & $-3.80$   &     $-3.80$   & --        \\
$1.20$       &  $10.90$   & 10.94  & $11.45$  & $-3.93$   &     $-3.93$   & --        \\
$1.10$       &  $14.52$   & 14.56  & $15.26$  & $-4.04$   &     $-4.04$   & --        \\
$1.00$       &  $17.99$   & 18.06  & $19.08$  & $-4.12$   &     $-4.12$   & --        \\
\hline
\end{tabular}
\label{table:Ekman_sigmaI_betaI}
\tablefoot{Coefficients obtained by fitting the DNS to the cubic Stuart-Landau model and directly from WNL theory (the amplitude and the multiscale expansions). Note that, in multiscale expansion method, the second Landau coefficient is independent of the Ekman number and the value reported is evaluated at $E_\mathrm{c}$ according to Eq. (\ref{eqn:sigma_beta}).}
\end{table*}

\section{Transport of angular momentum by the most unstable mode} \label{app:angular_momentum}

The second-order equation (\ref{eqn:2nd_order1a}) that governs the evolution of the mean flow $\psi_{20}$ can be reorganized in a form showing the transport of angular momentum. Equation~\eqref{eqn:2nd_order1a} can be written as
\begin{equation}
    \left[\left(\frac{2\sigma_{I}}{\Omega_\mathrm{ref}} - 2E\right)\mathbb{L}_0 - E \mathbb{L}_0^2\right]\psi_{20} = -\frac{2 m}{ \sin\theta} \Im \left[\frac{\mathrm{d}}{\mathrm{d}\theta} \left( \psi_{11} \overline{\zeta_{11}} \right) \right] ,
\label{eqn:2nd_order_rew}
\end{equation}
where $\mathbb{L}_0$ is the Legendre operator Eq.~\eqref{eqn:L_A} with $m=0$. Integrating the above equation with respect to  $\theta$ gives (after algebraic manipulations):
\begin{align}
    \frac{\dd}{\dd \theta}\left\{\left(\frac{2\sigma_{I}}{\Omega_\mathrm{ref}} - 2E\right)\psi_{20} - E \mathbb{L}_0 \psi_{20}\right\} = -\frac{2m}{\sin \theta} \Im \left(\psi_{11}\overline{\zeta_{11}}\right) . 
\label{eqn:2nd_order_rew1}
\end{align}
By using the velocity-stream function relation (\ref{eqn:straamfun1}), and the continuity equation for the velocity field $\mathbf{u}_{11}$, it can be shown that the RHS of Eq. (\ref{eqn:2nd_order1a}) is related to $Q_{11}^{\theta\phi} := \langle u_{11}^\theta u_{11}^\phi\rangle_\phi$, the Reynolds stresses between $u_{11}^\theta$ and $u_{11}^\phi$, through the relation
\begin{equation}
    2m \Im \left(\psi_{11}(\theta)\overline{\zeta_{11}(\theta)}\right) = -\frac{1}{r^2 \Omega_\mathrm{ref}^2}\frac{1}{\sin \theta}\frac{\dd}{\dd \theta}\left(\sin^2 \theta Q_{11}^{\theta\phi} \right) . 
\label{eqn:rhs1}
\end{equation}
Substituting Eq.~\eqref{eqn:rhs1} onto Eq.~\eqref{eqn:2nd_order1a} and integrating one more time  over $\theta$ leads to
\begin{equation}
    2\sigma_I r^2 \sin^3 \theta \Omega_{20} - \nu \frac{\partial}{\partial\theta}\left(\sin^3 \theta \frac{\partial \Omega_{20}}{\partial \theta}\right) + \frac{\partial}{\partial \theta}\left(\sin^2 \theta Q_{11}^{\theta\phi} \right) = 0 .
\label{eqn:ang_tr}
\end{equation}
As $\sigma_I$ is coming from the time derivative, this equation is similar to the transport of angular momentum equation \citep[see for example Sect.~4.1 in][]{Rudiger1989}. 
Hence the second-order equation of the WNL theory describes the transport of angular momentum by the most unstable mode. Please note that, in the multiscale expansion, we solve the linear eigenvalue problem at the critical Ekman number. Thus $\nu = \nu_\mathrm{c} = r^2\Omega_\mathrm{ref} \Ec$ which implies $\sigma_I = 0$ in Eq. (\ref{eqn:ang_tr}), and hence in that case we only get the large time limit for the relation of Reynolds stresses and the change in differential rotation, but not its time evolution.

\section{Frequency change of the fundamental from Stuart-Landau equation} \label{app:time_evolution}
The cubic Stuart-Landau equation is given by (\ref{eqn:cubic_SL_eqn1}) 
\begin{equation}
    \frac{\mathrm{d} A}{\mathrm{d}t} = -i \sigma A - i \beta |A|^2 A \, . \label{eqn:cubic_SL}
\end{equation}
Writing the complex time-dependent amplitude $A(t)$ as $A(t) = |A(t)|\rm{e}^{-\ii\varphi(t)}$, we obtain two equations for the amplitude $|A|$ and the phase $\varphi$
\begin{align}
    &\frac{\mathrm{d}|A|^2}{\mathrm{d}t} = 2\sigma_i |A|^2 + 2 \beta_I |A|^4 \, , \label{eqn:amp111} \\
    &\frac{\mathrm{d} \varphi}{\mathrm{d}t} = \sigma_R + \beta_R |A|^2 \, . \label{eqn:phase111}
\end{align}
There are two steady solutions to Eq. (\ref{eqn:amp111}):  the trivial equilibrium solution $|A|_\mathrm{eq} = 0$ and $|A|_\mathrm{eq} = \sqrt{-\sigma_I/\beta_I}$ providing that $\beta_I < 0$, which is the case for the the supercritical branch.

Since the temporal rate of change of local phase gives the instantaneous frequency, the frequency in the saturation regime can be inferred directly from the phase equation (\ref{eqn:phase111})
\begin{equation}
    \omega_\mathrm{eq} = \sigma_R + \beta_R |A|_{\rm eq}^2 \,.
\end{equation}
The linear eigenfrequency is thus modified by $\beta_R |A|_{\rm eq}^2 = -\beta_R \sigma_I / \beta_I$.

Using these limiting expressions for the amplitude, we write the fundamental mode in the saturated state ($t \to \infty$) as
\begin{align}
    \Psi(t, \theta, \phi) \to & |A|_\mathrm{eq} \Psi_{11}(\theta) \rm{e}^{-\ii\left(\omega_\mathrm{eq} t -m\phi\right)} +\mathrm{c.c.} + \mathcal{O}(|A|_\mathrm{eq}^2) \, . \label{eqn:limit_cycle_exp}
\end{align}
This epxression is accurate in $|A|_\mathrm{eq}$ but can be easily extended to $\mathcal{O}(|A|_\mathrm{eq}^3)$ by including the base flow modification, fundamental mode correction, second and third harmonics in Eq. (\ref{eqn:limit_cycle_exp}).

\end{document}